\documentclass{article}

     \usepackage[final]{neurips_2018}

\usepackage[utf8]{inputenc} 
\usepackage[T1]{fontenc}    
\usepackage{hyperref}       
\usepackage{url}            
\usepackage{booktabs}       
\usepackage{amsfonts}       
\usepackage{nicefrac}       
\usepackage{microtype}      

\usepackage{times}
\usepackage{epstopdf}
\usepackage{helvet}
\usepackage{courier}
\usepackage{caption}
\usepackage{graphicx}
\usepackage{tabularx}
\usepackage{amssymb,amsmath}
\usepackage{latexsym}
\graphicspath{ {plots/} }
\usepackage[lined,boxed,commentsnumbered]{algorithm2e}

\newcommand{\mmp}[1]{}
\newcommand{\yw}[1]{}
\newcommand{\yaliwan}[1]{}
\newcommand{\comment}[1]{}
\newcommand{\techreport}[1]{}
\newcommand{\myemph}[1]{{\em #1}}
\newcommand{\mydef}[1]{{\em #1}}
\newcommand{\beq}{\begin{equation}}
\newcommand{\eeq}{\end{equation}}
\newcommand{\beqa}{\begin{eqnarray}}
\newcommand{\eeqa}{\end{eqnarray}}
\newcommand{\bit}{\begin{itemize}}
\newcommand{\eit}{\end{itemize}}
\newcommand{\benum}{\begin{enumerate}}
\newcommand{\eenum}{\end{enumerate}}

\newtheorem{prop}{Proposition}
\newlength{\backitem}
\newsavebox{\savestuff}

\newcommand{\G}{{\cal G}}

\newcommand{\fracpartial}[2]{\frac{\partial #1}{\partial  #2}}

\newcommand{\nodes}{{\mathcal V}}

\newcounter{Examplecount}
\newcommand{\distance}{\operatorname{dist}}

\newcommand{\rrr}{{\mathbb R}}

\newcommand{\clust}{\mathcal{C}}

\newcommand{\IF}{$I\!F$}
\newcommand{\tilda}{\tilde{A}}
\newcommand{\tilddd}{\tilde{d}}
\newcommand{\tildl}{\tilde{L}}
\newcommand{\tildd}{\tilde{D}}
\newcommand{\BP}{$BP$}

\newcommand{\Gw}{\tilde{\G}}

\title{Measuring the Robustness of Graph Properties}

\author{%
  Yali Wan \\
  Amazon.com\\
  Palo Alto, CA 94301 \\
  \texttt{yalwan@amazon.com} \\
\And
Marina Meila\\
University of Washington \\
Seattle, WA 98195\\
 \texttt{mmp@stat.washington.edu} \\
}

\begin{document}

\maketitle

\begin{abstract}
In this paper, we propose a perturbation framework to measure the robustness of graph properties. Although there are already perturbation methods proposed to tackle this problem, they are limited by the fact that the strength of the perturbation cannot be well controlled. We firstly provide a perturbation framework on graphs by introducing weights on the nodes, of which the magnitude of perturbation can be easily controlled through the variance of the weights. Meanwhile, the topology of the graphs are also preserved to avoid uncontrollable strength in the perturbation. We then extend the measure of robustness in the robust statistics literature to the graph properties. 
\end{abstract}

\section{Introduction}\label{sec:intro}
In data science it is natural to consider an observed graph, or network, $\G$ as a
realization of a random process. Consequently, the properties of $\G$
 (such as diameter, conductance, clustering)  and the inferences drawn
from them are incomplete without some measure of variability or confidence. \\

We propose a new methodology for evaluating the robustness of graph
properties by measuring the effect of small
perturbations of the graph on the respective property. This methodology is based on
concepts from robust statistics \cite{Hampel:86} and exploits the
technique introduced in \cite{MPentney:sdm07} to augment
a graph Laplacian with user-defined weights.  Let $\G=(\nodes,A)$ be a
graph with $n=|\nodes|$ nodes and adjacency matrix $A$; $A$ can be
symmetric or asymmetric (corresponding to a directed network), and we
assume $A_{ij}\in\{0,1\},\,A_{ii}=0$ (simple graph) or $A_{ij}\geq 0$
(weighted graph). Our methodology applies to both scenarios, and
makes no assumptions on how the graph was generated.\\

Our methodology consists of four key components:
\benum
\item \myemph{Perturb} the nodes of the graph by assigning them multiplicative weights, with $w_i$ the weight of node $i$, $i=1,2,\ldots n$. If $w_i=1$ for all $i$, we have the original graph $\G$. 
\item Express the desired graph property $f(\G)$ as a \myemph{smooth function of the weights}.
\item Construct \myemph{measures of robustness} inspired by the {robust statistics} literature, such as \myemph{influence function (\IF)}, \comment{\myemph{gross error sensitivity}} and \myemph{breakdown point (\BP)}.
\item \myemph{Evaluate} these measures on the current graph $\G$. 
\eenum
We exemplify our approach by examining the robustness of weighted cut (WCut), number of weakly Connected Components (wCC's), eigengap and clustering. In this paper, we make use of the following notations. We define $w_i, i=1,2, \cdots, n$, the weight associated with node $i$. Since we are interested in both directed and undirected graphs,  we define $d_i = \sum_{j=1}^nA_{ij}$ as the out-degree of node $i$. We use the definition in \cite{MPentney:sdm07} and define $L  = I - \frac{1}{2}D^{-1/2}(A + A^T)D^{-1/2}$ as the Laplacian matrix associated with $A$. This definition is consistent with the usual definition of $L = I - D^{-1/2}AD^{-1/2}$ when the graph is symmetric. We define $A$, $d$, $D$, $L$ as the properties of the observed graph. The above quantities will be marked with a symbol $\tilde{}$ when they are perturbed. This is consistent with the notation in matrix perturbation literature. To demonstrate, $\tilda$ represents the adjacency matrix of the perturbed graph.\\ 


In the rest of this paper, we proceed as follows. In Section \ref{sec:2}, we describe our method of bootstrapping and perturbing the networks. In Section \ref{sec:properties}, we talk about the graph properties of interest. In Section \ref{sec:if}, we discuss measures for evaluating robustness of graph properties. In Section $5$, we discuss breakdown points. In Section \ref{sec:related}, we discuss related work. In Section \ref{sec:5}, we use both synthetic and real datasets to analyze our methods. We conclude our findings in Section \ref{sec:6}.\\

\section{Perturbing the network} \label{sec:2}
Existing methods of perturbing networks can be found in recent works \cite{karrerLN:08, gfeller2005finding, bhattacharyya2015subsampling, ali2016comparison}.  They mostly involve removing or duplicating edges or nodes randomly and independently in the graph. \cite{gfeller2005finding} randomly removes edges from the graph. \cite{karrerLN:08} maintains the the number of vertices and edges of the original graph, and perturbs the graph by moving the edges to the other locations.  On the other hand, \cite{bhattacharyya2015subsampling} and \cite{ali2016comparison} bootstrap the network by subsampling the nodes. Although their approach is straightforward, the perturbation cannot be well controlled. Firstly, the topology of the graphs can change dramatically for sparse graphs. Randomly removing or adding the nodes and edges makes the strength of perturbation hard to control. For example, assume we have two densely clusters $\clust_1$ and $\clust_2$ connected by a single edge $e$. The effect of removing $e$ versus removing an edge within $\clust_1$ or $\clust_2$ is very different, since the former will change the graph structure drastically by making it disconnected. Moreover, removing or adding edges (nodes) are discrete moves, therefore the perturbation cannot be arbitrarily small.\\

In order to have a fine control on the amount of perturbation and make it as smooth as possible, in our approach, we preserve the graph topology by keeping all the nodes and edges, and we perturb the graph by assigning random weights to the nodes and then distribute these weights to edges. Specifically, we assign weight $w_i$ to node $i$, where $w_i$ is generated i.i.d from a distribution with $E(w_i) = 1$,  standard deviation $\sigma_w$, and support on $(0, \infty)$. The perturbation can be controlled smoothly with $\sigma_w$.
Since $w_i > 0$, no nodes or edges is removed or added, and the topology of the graph is preserved.\\

We propose two ways of constructing $\tilda$ from weighted nodes.
\bit
\item Asymmetric perturbation: perturb outgoing edges of node $i$, so that $\tilda_{ij} = w_iA_{ij}$ (whereas $\tilda_{ji} = w_jA_{ji}$). The out-degree becomes $\tilddd_i = d_i w_i$. The perturbed Laplacian is,
\beq\label{eq:lhata}
\tildl_{ij} = 1 - \frac{\tilda_{ij} + \tilda_{ji}}{2\sqrt{\tilddd_i \tilddd_j}} = 1 - \frac{w_iA_{ij} + w_jA_{ji}}{2\sqrt{w_iw_jd_id_j}}
\eeq
\item Symmetric perturbation: place $w_i$ on both outgoing edges and incoming edges of node $i$.  $\tilda_{ij} = (w_i + w_j - 1)A_{ij}$. Since this method leads to complicated $\tilde{L}$ in form, we discuss perturbing one node at a time. If node $t$ is perturbed, the Laplacian becomes,
\beq\label{eq:lhat1}
\tildl_{ij} = \begin{cases}
		1 - \frac{A_{tj}w_t + A_{jt}w_t}{2\sqrt{d_tw_t(\sum_{m\neq t} A_{jm} + A_{jt}w_t)}} & i=t, j\neq t \\
		1 - \frac{A_{ij} + A_{ji}}{2\sqrt{(\sum_{m \neq t} A_{jm} + A_{jt}w_t)(\sum_{m \neq t}A_{im} + A_{it}w_t)}} & i,j \neq t \\
		 0 & i, j = t\\
		\end{cases}
\eeq
\eit
Both of the above methods ensure $E(\tilda_{ij}) = A_{ij}$. Notice that in the current methods, it can be easily shown that bias is introduced in $A_{ij}$, in which case one cannot separate the perturbation from structural change and the change of values in adjacency matrix.\\

 Although both perturbation methods seem to be reasonable, they have graph-specific advantages. The asymmetric method provides a very simple way for calculating a graph property like $\tilddd$ after perturbation. On the other hand, it is not interesting for the perturbation of weakly Connected Components (wCC's), as we shall see in Section \ref{sec:if}. On the other hand, the symmetric method maintains a symmetric perturbation of adjacency matrix. If $A_{ij} = A_{ji}$, then $\tilda_{ij} = \tilda_{ji}$ after perturbation. However, $\tildl$ is complicated in form. The symmetric perturbation is not very interpretable for some graph properties (WCut), but it is useful for evaluating the robustness in eigengap and wCC's.\\

In the above methods, we put weights on nodes and then distribute the weights to edges. The reason for doing this instead of directly perturbing the edges is that it is more natural to maintain the association between nodes and edges. The edges connected to the same node should be dependent rather than independent. Perturbing the weights on edges \cite{karrerLN:08} independently neglects these relationships, which has been found both unrealistic and disrupts the topology of sparse network. Moreover, in many applications, nodes are more meaningful than edges. Firstly, a node often carries its own attributes. Secondly, a node is described by the multiple nodes it is connected to. On the other hand, it is more difficult to glean information from an edge. For example, in a Facebook network, a node is a person with complex information including age, birth place, school, the friends he connected to, etc. While an edge is formed when two people befriend each other, and we cannot even learn how well these two people know each other and it tells us little on other information.\\

\subsection{The bias in $\tildl$}
The graph properties we shall study are closely related to $L$. It would be nice if the changes in graph properties are only from the noise introduced in graph structure and keeping the entires in $\tildl$ to be the same with that in $L$ in expectation. Since having bias in $L$ introduces another factor for the change in graph property, of which the strength of perturbation is hard to control. Therefore in principle, we would want $E(\tildl_{ij}) = L_{ij}$. In the following paragraphs we prove that this is impossible under all the current perturbation methods. For the current methods, where people move the edges and subsample the nodes, the bias in $\tildl$ is apparent. We show in Proposition \ref{prop:1} that both of our perturbation methods introduce bias too.\\
\begin{prop}\label{prop:1}
Assume $w_i, i=1:n$ are generated i.i.d from a distribution with $E(w_i) = 1$, $\sigma_w\neq 0$, $w_i > 0$. For asymmetric perturbation, $E(\tildl_{ij}) < L_{ij}$.\\
If we further assume there exists a triangle $<t, p, q>$ in the graph. For symmetric perturbation, with node $t$ perturbed, $E(\tildl_{ij}) < L_{ij}$.
\end{prop}
{\bf Proof:} \\
In the asymmetric perturbation, $\tildl$ is shown in Equation \ref{eq:lhata}. Assume $L_{ij} \neq 0$. In order to have $E(\tildl_{ij}) = L_{ij}$, we need
\beq
E(1 - \frac{w_iA_{ij} + w_jA_{ji}}{2\sqrt{w_iw_jd_id_j}}) = 1 - \frac{A_{ij} + A_{ji}}{2\sqrt{d_id_j}}
\eeq
\beq
A_{ij}(1 - E(\sqrt{\frac{w_i}{w_j}})) + A_{ji}(1 -E(\sqrt{\frac{w_j}{w_i}})) = 0
\eeq
Since $w_i$ and $w_j$ are i.i.d, we then have,
\beq
E(\sqrt{\frac{w_i}{w_j}}) = E(\sqrt{\frac{w_j}{w_i}}) = E(\sqrt{w_i}) E(\frac{1}{\sqrt{w_j}}) =  E(\sqrt{w_i}) E(\frac{1}{\sqrt{w_i}}) = 1
\eeq
Assume $x = \sqrt{w_i}$. Equivalently we have, 
\beq
E(\frac{1}{x}) = \frac{1}{E(x)}
\eeq
Assume $f(x) = \frac{1}{x}$. Since $f(x)$ is strictly convex and Jensen's inequality yields, $E(f(x)) \ge f(E(x))$, that is,
$E(\frac{1}{x}) \ge \frac{1}{E(x)}$. where equality holds only when $\sigma(x) = 0$.  Therefore $E(\sqrt{\frac{w_i}{w_j}}) > 1$. We have reached a contradiction. $E(\tildl) < L$.\\
In the symmetric perturbation, $\tildl$ after perturbing node $t$ is shown in Equation \ref{eq:lhat1}. In order have $E(\tildl_{ij}) = L_{ij}$, we need,
\beq
E(1 - \frac{A_{tj}w_t + A_{jt}w_t}{2\sqrt{d_tw_t(\sum_{m\neq t} A_{jm} + A_{jt}w_t)}}) = 1 - \frac{A_{tj} + A_{jt}}{2\sqrt{d_td_j}} \text{ when }i=t, j\neq t
\eeq
\beq
E(1 - \frac{A_{ij} + A_{ji}}{2\sqrt{(\sum_{m \neq t} A_{jm} + A_{jt}w_t)(\sum_{m \neq t}A_{im} + A_{it}w_t)}}) = 1 - \frac{A_{ij} + A_{ji}}{2\sqrt{d_id_j}}  \text{ when } i, j \neq t 
\eeq
That is, 
\beq\label{eq:p11}
E(\sqrt{w_t})E(\frac{1}{\sqrt{a_j + (1-a_j)w_j}}) = 1,  \text{ when }i=t, j\neq t,
\eeq
\beq\label{eq:p12}
E(\frac{1}{\sqrt{a_i + (1-a_i)w_i}})E(\frac{1}{\sqrt{a_j + (1-a_j)w_j}}) = 1, \text{ when } i, j \neq t,
\eeq
where $a_i = \frac{\sum_{m \neq t} A_{jm}}{d_j}$ and $0\le a_i \le 1$. \\
Since there is a triangle $<t, p, q>$ in the graph. We can easily derive from Equation \ref{eq:p11} and \ref{eq:p12} that $E(\sqrt{w_t}) = 1$, and $E(\frac{1}{\sqrt{a_i + (1-a_i)w_i}}) = 1$, $\forall i$.\\
$E(\frac{1}{\sqrt{a_i + (1-a_i)w_i}})$ is a continous and differentiable function of $a_i$. When $a_i = 1$, $E(\frac{1}{\sqrt{a_i + (1-a_i)w_i}}) = 1$. When $a_i = 0$, $E(\frac{1}{\sqrt{a_i + (1-a_i)w_i}}) = E(\frac{1}{\sqrt{w_i}}) > \frac{1}{E(\sqrt{w_i})} = 1$. Since $\frac{\partial}{\partial a_i}E(\frac{1}{\sqrt{a_i + (1-a_i)w_i}}) < 0$,  $E(\frac{1}{\sqrt{a_i + (1-a_i)w_i}}) > 1$ for $0 < a_i < 1$. We have a contradiction. Therefore in conclusion,  $E(\tildl)< L$.$\blacksquare$

We have proved that in theory there will be bias introduced in $\tildl$, it would be a good idea to examine how much bias will actually be introduced in practice. Under asymmetric perturbation, one can easily derive from the proof in Proposition \ref{prop:1} that $E(\sqrt{w})E(\frac{1}{\sqrt{w}}) = \frac{\tildl_{ij}}{L_{ij}}$. Therefore, the more $E(\sqrt{w})E(\frac{1}{\sqrt{w}})$ deviates from $1$, the more bias introduced in $\tildl_{ij}$. Since there are very limited number of conventional distributions that allow support on $(0, \infty)$ and mean equals $1$, we propose a class of mixture distributions $Mixture(a, b, \sigma_-^2, \sigma_+^2, T_-, T_+, p)$, where $ap + b(1-p) = 1$, $T_-$ is a distribution with $E(x) = a$, $\sigma(x) = \sigma_-^2$, with support on $(0, 1)$, $T_+$ is a distribution with $E(x)=b$, $\sigma(x) = \sigma_+^2$, with support on $(1, \infty)$. Assume $w\sim Mixture(a, b, \sigma_-^2, \sigma_+^2, T_-, T_+, p)$, then $p(w\sim T_-) = p$, $p(w\sim T_+) = 1-p$. We can easily show that $E(w) = 1$, $\sigma^2(w) = p_-\sigma_-^2 + p_+\sigma_+^2$. The mixture distribution subsumes all the distributions concentrated on $1$, with support on $(0, \infty)$. \\

\begin{figure}[ht]
\centering
\includegraphics[width=0.75\linewidth]{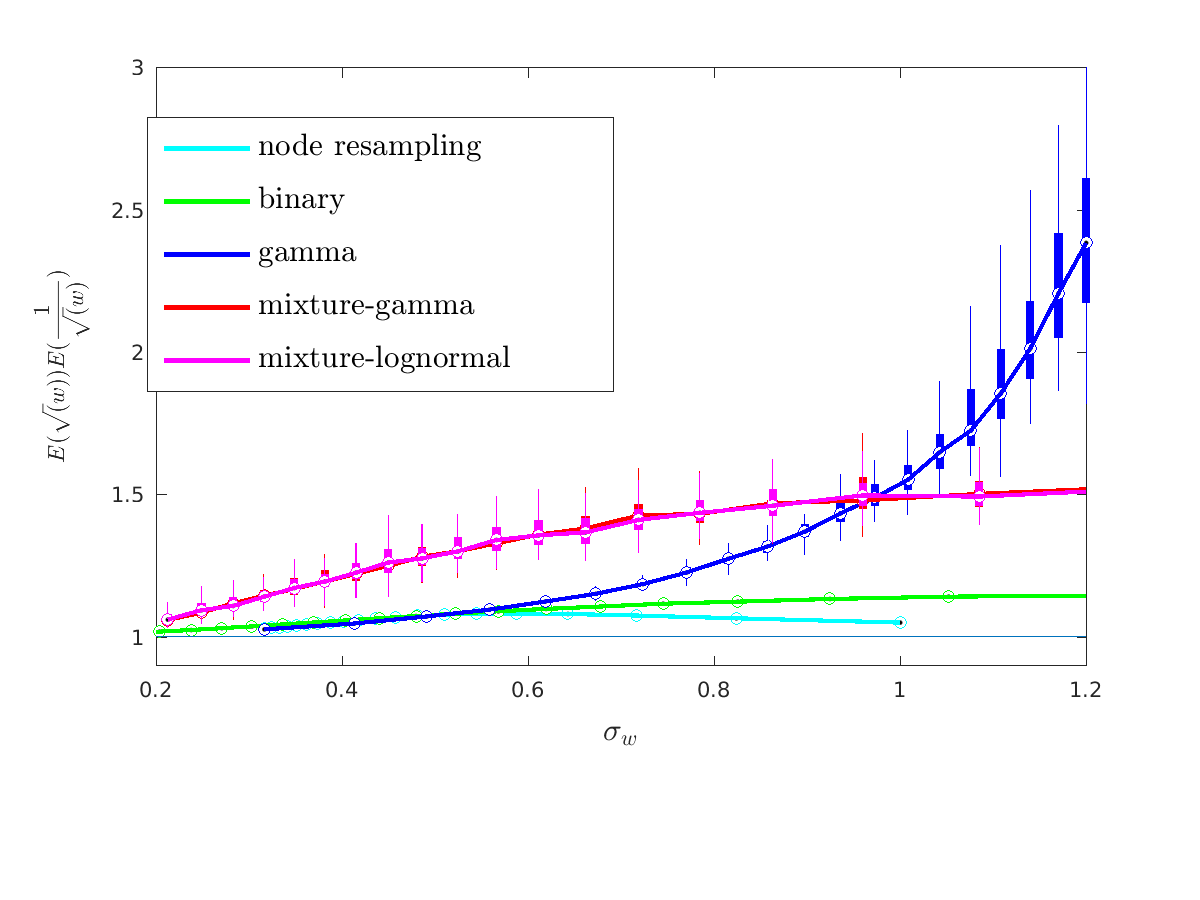}\\
\caption{\label{fig:bias}\label{fig:w}$E(\sqrt{w})E(\frac{1}{\sqrt{w}})$ under asymmetric perturbation. Each boxplot represents 100 repetitions. }
\end{figure}

In the following experiment, we generate $w_i$ from the distributions below accordingly.
\benum
\item Node resampling: $w_i$ is obtained from resampling the nodes with replacement to form a sample of size $N$.  Assume node $i$ appears $m$ times in the sample, $w_i = \frac{mn}{N}$. $\sigma_w = \sqrt{\frac{n-1}{N}}$. We can easily see as $N$ goes up, $\sigma_w$ decreases. We can then control $\sigma_w$ by varying the size of $N$.
\item Binary distribution: $w_i$ follows binary distribution on $\{a, b\}$, with $P(w_i = a) = p$, $P(w_i = b) = 1-p$. $b = \frac{1-ap}{1-p}$. $E(w_i) = 1$, $\sigma_w^2 = p(a-1)^2 + (1-p)(b-1)^2$.
\item Gamma distribution: $w_i \sim Gamma(a, \frac{1}{a})$. $E(w_i) = 1$, $\sigma_w^2 = \frac{1}{a}$. 
\item Mixture-Gamma-Uniform distrbution: $a = 0.5$, $T_- = uniform(0,1)$, $T_+ = Gamma(\frac{b-1}{\sigma_+}, \sigma_+) + 1$.
\item Mixture-Lognormal-Uniform distribution: $a=0.5$, $T_-=uniform(0,1)$, $T_+ = lognormal(b-1, \sigma_+)$.
\eenum
The results are shown in Figure \ref{fig:bias}. We observe that the bias is sensitive to the choice of weight distribution. In specific, the bias introduced by Gamma distribution is growing exponentially with $\sigma_w$, thus not a good option. Node resampling and Binary distribution generate smallest bias. However node resampling can only allow the perturbation strength $\sigma_w$ to be as much as $1$, otherwise the topology of the graph is changed. Binary distribution only allows two choices of weight, which is very limited. The mixture distributions behave similarly in terms of introducing bias. Although the bias is nontrivial, the fact that it is upper bounded and that $w_i$ can be generated in a continuous manner make them good candidates for performing perturbations on graphs.\\

We also evaluate the bias introduced in $\tildl$ in the current methods including \cite{karrerLN:08} and \cite{bhattacharyya2015subsampling}. Since they move the edges or delete nodes, the bias depends on the graph. We generate a graph from DC-SBM model \cite{qin2013regularized} with $n=800$, $w_i \sim 0.5 + uniform(0.5, 1)$, and evaluate $\frac{E(\tildl_{ij})}{E(L_{ij})}$. As a reminder, $\frac{E(\tildl_{ij})}{E(L_{ij})} = E(\sqrt{w})E(\frac{1}{\sqrt{w}})$. 
\begin{figure}[ht]
\begin{tabular}[b]{cc}
\centering
\includegraphics[width=0.5\linewidth]{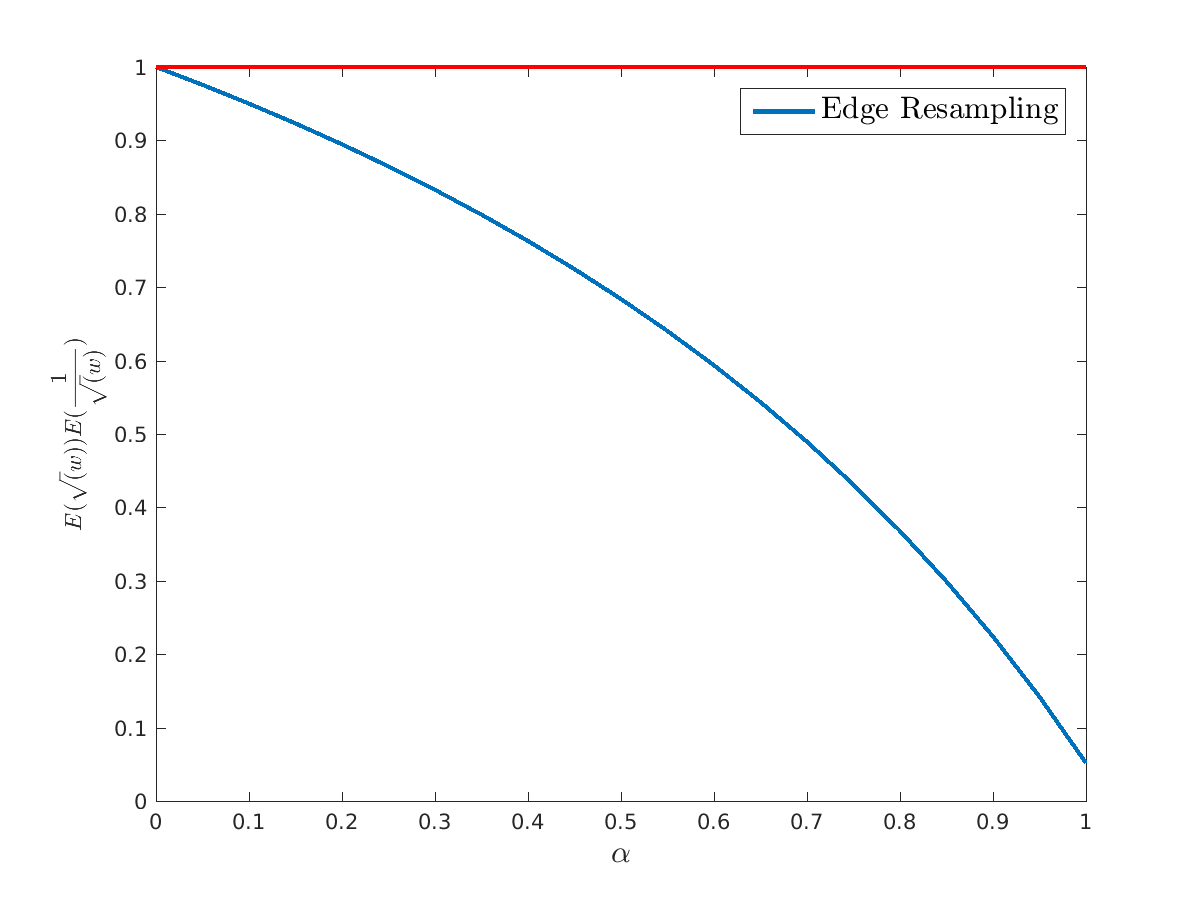}&
\includegraphics[width=0.5\linewidth]{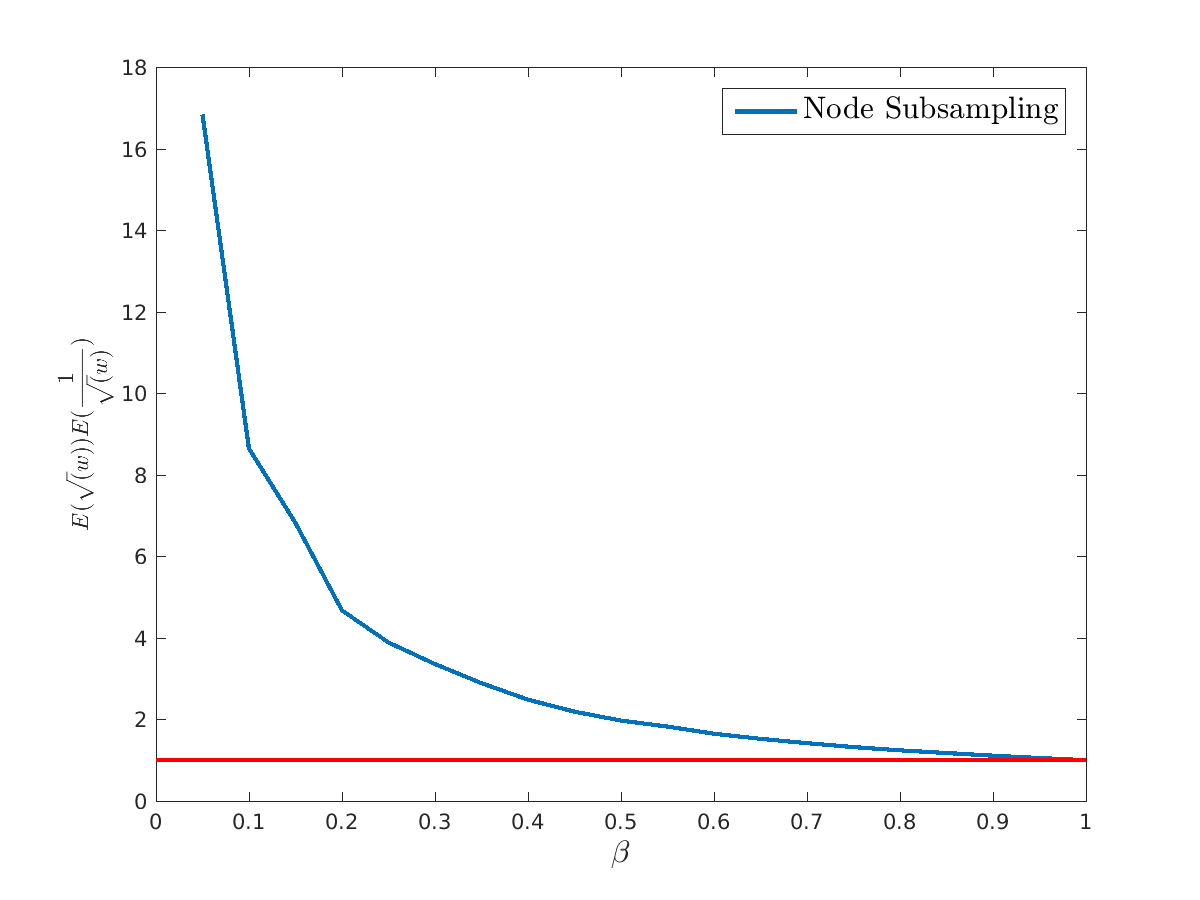}\\
perturbation strength & proportion of nodes sampled\\
\end{tabular}
\caption{\label{fig:biasrelated} Left: the bias from the method in \cite{karrerLN:08}. $\alpha$ indicates the strength of perturbation. Right: the bias from the method in \cite{bhattacharyya2015subsampling}. $\beta$ indicates the proprotion of nodes that are subsampled. Larger $\beta$ indicates larger perturbation strength.}
\end{figure}
In Figure \ref{fig:biasrelated}, we show that both methods introduce large bias compare to our perturbation methods.\\

\subsection{Partial perturbation and full perturbation}
In this section we discussed perturbing the graph by adding i.i.d weights to the nodes. Although we make i.i.d assumption on $w_i$, it can be relaxed. In this paper, we consider two approaches to utilize the weight perturbation.  First, in order to evaluate the sensitivity of graph properties, we perturb all the nodes i.i.d with different perturbation strength $\sigma_w$, and then investigate how the graph property changes with respect to it. In specific, we fix $E(w)=1$ and vary $\sigma_w$ for all the nodes. Alternatively, in order to probe the source of sensitivity, we can perturb a subset of nodes. For these nodes, we perturb their weights i.i.d by fixing $\sigma_w$ and vary $E(w)$, from which we can discover the source of the robustness in graph properties. The use of these two perturbation approaches will be shown in Section \ref{sec:5}.\\

\section{Expressing graph properties with weights} \label{sec:properties}
The success of our approachs depend on our ability to express properties of interest as functions $f(\G,w)$, which are continuous and differentiable w.r.t. the node weights $w_{1:n}$. In this project we focus on graph properties that depend on graph Laplacians. Many important graph properties depend on Laplacians. For instance, the \myemph{mixing time} of the graph depends on the second smallest eigenvalue $\lambda_2(L)$. Diffusion distances \cite{nadler:06} between nodes can also be approximated by the principal eigenvectors and values of $L$. The number of connected components $C$ of $\G$ is equal to the multiplicity of 0 in the spectrum of $L$, etc. There are four graph properties that we are particularly interested in, which will be used as examples in the following sections: the weighted cut of the graph (WCut), the number of weakly connected components, the eigengap and clustering.\\

\subsection{Weighted Cut(WCut)} 
Weighted cut is defined as a graph property associated with clustering in the general graph setting \cite{MPentney:sdm07}, which can be used for both directed and undirected graphs. It is similar in motivation to the normalized cut for undirected graph. Both aim in finding a cut of low weight in the graph while balancing the sizes of the clusters.  The multiway version of the normalized cut MNCut of \cite{MPentney:sdm07} is a special case of WCut. \\

Formally, WCut with respect to clustering $\clust$ with $K$ clusters is defined as
\beq
WCut(\G, w, \clust) = \sum_{k=1}^K \frac{1}{\tildd_k}\sum_{i\in \clust_k}(\tilddd_i - \sum_{j \in \clust_k} \tilda_{ij})
\eeq

where $\tildd_k = \sum_{i \in \clust_k} \tilddd_i$. Further define $\tildd_{kk} = \sum_{i \in \clust_k}\sum_{j\in \clust_k} \tilda_{ij}$, we can then equivalently write Wcut as,
\beq
WCut(\G, w, \clust) = \sum_{k=1}^K (1 - \frac{\tildd_{kk}}{\tildd_k})
\eeq

Small WCut suggests sparse connections between clusters, thus better quality of clustering.
\subsection{Number of weakly Connected Components (wCC's) and eigengap} \label{subsec:cc}
It is already know that the number of connected components (CC's) is not robust, since randomly adding a node or removing some edges can easily change the number of CC's. Instead, we study the number of weakly Connected Components (wCC's), where ``weakly'' means sparse connections between CC's. \\

We propose a pair of functions $f_u(\G,w, K)=\lambda_{K+1}(L(\Gw))-\sum_{k=1}^K\lambda_i(L(\Gw))$, $f_l(\G,w, K)=\sum_{k=1}^K\lambda_k(L(\Gw))$ to describe the number of wCC's. The reason we use $f_l$ is that, if there are $K$ number of CC's, then $f_l = 0$. We would then expect $f_l$ to be close to $0$ for $K$ number of wCC's. We choose $f_u$ because we expect a significant gap between $f_l$ and $\lambda_{K+1}(L(\Gw))$ for a stable number of wCC's. When $f_u$ is away from 0 (respectively $f_l$ near 0) there are exactly $K$ ``weakly connected'' components in $\Gw$. If this holds for large perturbations, then $K$ can be considered robust. \\

The $K$th eigengap is defined as $f_e(\Gw, w, K) = \lambda_{K+1}(L(\Gw)) - \lambda_K(L(\Gw))$. It indicates $K$ principal subspace when $f_e$ is large compared to the other eigengaps.\\

Notice that function $f$ defined for number of wCC's and eigengap are only meaningful for undirected graphs. Since the eigenvalues is not very interpretable for the directed graphs.

\section{Influence functions} \label{sec:if}
In order to evaluate the graph properties, we construct the target properties as differentiable functions $f(\G, w)$ and see how much $f(\G, w)$ changes with respect to the size of perturbation of $w$. In this section, we present tools for quantifying robustnes including \myemph{Influence
  Function} (IF) and \myemph{Breakdown Points}. We firstly talk about Influence function (IF), which was invented by Hampel in \cite{Hampel:86}. The importance of IF lies in its interpretation: it gives a picture of the infinitesimal behavior of the asymptotic value. While numerous studies have been done on methods for analysis of data sampled from a known distribution i.i.d, there has not been much work on using these tools to evaluate the robustness of graph properties. Here we define for a perturbed graph $\Gw$,
\beq
IF_t=\left.\fracpartial{f(\G,w)}{w_t}\right|_{w_{1:n}=1}, 
\eeq
which measures the local influence of $w_t$ on $f$.\\

\subsection{Influence function of WCut}

\begin{prop}
Assume a graph $\G$ with $\clust$, $d_i$ defined as usual. Assume node $t \in \clust_{k_0}$. $d_{ik} = \sum_{j\in\clust_k}A_{ij}$,  $d_{ki} = \sum_{j \in \clust_k}A_{ji}$. $D_{k_0k_0^{\neg}} = \sum_{i
\in\clust_{k_0},j \notin \clust_{k_0}} A_{ij}$. Then using asymmetric perturbation, the influence function for node $t$ is 
\beq
I\!F^{WCut}_t(\G, \clust) =\frac{d_tD_{kk} - d_{tk}D_{k}} {D_k^2}.
\eeq
 \\
Using symmetric perturbation. 
\beq
I\!F^{WCut}_t(\G, \clust) =  \sum_{k=1}^n \frac{D_{kk} d_{kt}}{D_{k}^2} - \frac{D_{k_0}d_{k_0t} + D_{k_0k_0^{\neg}}D_{tk_0} - D_{k_0k_0}\sum_{j\notin\clust_{k_0}}A_{tj}}{D_{k_0}^2}
\eeq
\end{prop}
{\bf Proof:}\\
We firstly perturb the graph using the asymmetric method. Assume $t \in \clust_k$, we have
\begin{align}
\frac{\partial WCut(\G, w)}{\partial w_t} &= \frac{\partial}{\partial w_t}\frac{1}{\sum_{j\in \clust_k} d_jw_j} \sum_{j\in \clust_k} w_j(d_j - d_{jk})\\
& = \frac{d_t\sum_{j \in \clust_k} w_j d_{jk} - d_{tk} \sum_{j\in \clust_k} d_j w_j} {(\sum_{j \in \clust_k} d_jw_j)^2}.
\end{align}

The influence function is then derived as
\begin{align}
\frac{\partial WCut(\G, w)}{\partial w_t}|_{w_{1:n} = 1} &= \frac{d_t\sum_{j \in \clust_k} d_{jk} - d_{tk} \sum_{j\in \clust_k} d_j} {(\sum_{j \in \clust_k} d_j)^2} = \frac{d_tD_{kk} - d_{tk}D_{k}} {D_k^2}\\
\end{align}

In the symmetric perturbation, since the perturbation can be explained as perturbing one node at a time, one can easily show that the influence function is same for perturbing one node or multiple nodes. For simplicity, we assume perturbing node $t$ with weight $w_t$. WCut can be written as
\beq
\begin{split}
WCut(\G, w) = \sum_{k, t\notin \clust_k} (1- \frac{D_{kk}}{\sum_{i\in C_k}[(\sum_{j\neq t}A_{ij}) + A_{it}w_t]}) + \\(1 - \frac{\sum_{i \in \clust_{k_0}}A_{it}w_t + \sum_{j \in \clust_{k_0}}A_{tj}w_t + \sum_{i, j\neq t, i, j \in \clust_{k_0}} A_{ij}}{d_tw_t + \sum_{i\in \clust_{k_0}, i \neq t}[(\sum_{j\neq t}A_{ij}) + A_{it}w_t]})
\end{split}
\eeq
\begin{align}
\frac{\partial WCut(\G, w)}{\partial w_t}|_{w_{1:n} = 1} &=  \sum_{k, t\notin \clust_k} \frac{D_{kk} \sum_{i\in\clust_k} A_{it}}{D_{k}^2} - \frac{(\sum_{i\in \clust_{k_0}}A_{it} + \sum_{j\in\clust_{k_0}}A_{tj})D_{k_0} - D_{k_0k_0}(d_t + \sum_{i\in\clust_{k_0}}A_{it})}{D_{k_0}^2}\\
& = \sum_{k=1}^n \frac{D_{kk} \sum_{i\in\clust_k} A_{it}}{D_{k}^2} - \frac{D_{k_0}\sum_{i\in\clust_{k_0}}A_{it} + D_{k_0k_0^{\neg}}\sum_{j\in\clust_{k_0}}A_{tj} - D_{k_0k_0}\sum_{j\notin\clust_{k_0}}A_{tj}}{D_{k_0}^2}\\
&=\sum_{k=1}^n \frac{D_{kk} d_{kt}}{D_{k}^2} - \frac{D_{k_0}d_{k_0t} + D_{k_0k_0^{\neg}}d_{tk_0} - D_{k_0k_0}\sum_{j\notin\clust_{k_0}}A_{tj}}{D_{k_0}^2}
\end{align}\\
$\blacksquare$

In the asymmetric perturbation, IF has an intuitive interpretation when a point has no influence, i.e, $IF_t = 0$, 
\beq
\frac{d_{tk}}{d_i} = \frac{D_{kk}}{D_k} = \frac{mean_{i\in\clust_k}{d_{ik}}}{mean_{i\in\clust_k}{d_i}}
\eeq
It means that, node $t$ has $0$ influence in WCut when the proportion of edges that goes to $\clust_K$ equals the cluster level ratio of averages. If $IF_t > 0$, node $t$ tends to make WCut larger when more weight is put upon $t$, the quality of clustering decreases since the clustering becomes less well separated. Node $t$ is therefore considered unstable to the clustering, or not well clustered. If $IF_t < 0$, node $t$ is well clustered since WCut will decrease if $t$ is weighted more. Hence IF measures the robustness of clustering in node level. It is worth noticing that the influence of a node depends only on the cluster that the node belongs to, and is independent of the rest of the clusters. Moreover, the influences of the nodes within a cluster always cancel each other. In a well separated clustering, we would expect the influences of the nodes the be concentrated around $1$. When the clusters are completed separated, that is, there is no edge between clusters, it can be easily shown that all the node influence equals $0$.\\

\beq
\sum_{i\in \clust_k}\frac{\partial WCut(\G, w)}{\partial w_i}|_{w_{1:n} = 1} = \frac{\sum_{i\in \clust_k}d_i\sum_{j\in\clust_k}d_{jk} - \sum_{i\in\clust_k}d_{ik}\sum_{j\in\clust_k}d_j}{(\sum_{j\in\clust_k}d_j)^2} = 0,
\eeq

For the symmetric perturbation, the meaning for the above results is not very interpretable, since its form is very complicated and $IF_t = 0$ does not provide us with any clear interpretation in its balance state. 

\subsection{Influence function of number of wCC's and eigengap}
Here we consider the property of wCC's and eigengap described by $f_u$, $f_l$ and $f_e$, which are proposed in Section \ref{subsec:cc}. Since they both depend on $\lambda$, we firstly study $\frac{\partial \lambda_k}{\partial w_t}$ for a single $\lambda_k$ in Proposition \ref{prop:3}.. The influence functions of interest can be easily derived from there. This is because the influence functions can be written as,
\beq
I\!F^{f_u}_t = \frac{\partial \lambda_{K+1}}{\partial w_t} - \sum_{i = 1}^K\frac{\partial \lambda_{i}}{\partial w_t}
\eeq
\beq
I\!F^{f_l}_t = \sum_{i = 1}^K\frac{\partial \lambda_{i}}{\partial w_t}
\eeq
\beq
I\!F^{f_e}_t = \frac{\partial \lambda_{K+1}}{\partial w_t} - \frac{\partial \lambda_{K}}{\partial w_t}
\eeq

Unfortunately, asymmetric perturbation is not very interesting for undirected graphs, since the properties remain untouched despite the perturbation, as will be shown in Proposition \ref{prop:4}. This is because the eigengap of $L$ is equal to the eigengap of the transition matrix $P=D^{-1}A$, and $P$ stays unchanged after the asymmetric perturbation. This motivates the use of symmetric perturbation, the results of which are shown in Proposition \ref{prop:3}.
\begin{prop}\label{prop:4}
Using asymmetric perturbation, assume $A$ symmetric and $\lambda_k$ of multiplicity $1$. $\frac{\partial \lambda_k}{\partial w_t}|_{w_{1:n} = 1}=0 $. $v_i$ is the $i$-th element of the $k$-th eigenvector of $L$.
\end{prop}
{\bf Proof:}
\beq \label{eq:lamIF}
\frac{\partial \tilde{\lambda}_k}{\partial w_t} = \sum_{ij}\frac{\partial \tilde{\lambda}_k}{\partial \tildl_{ij}}\frac{\partial \tildl_{ij}}{\partial w_t} = \sum_{ij} v_{i}v_{j}\frac{\partial \tildl_{ij}}{\partial w_t}
\eeq
Since $\tildl_{ij} = 1 - \frac{A_{ij}w_i + A_{ji}w_j}{2\sqrt{w_iw_jd_id_j}}$ We then obtain,
\beq
\frac{\partial \tildl_{ij}}{\partial w_i} =  \frac{-2A_{ij}\sqrt{w_iw_jd_id_j} + A_{ij}\sqrt{w_iw_jd_id_j} + A_{ji}\sqrt{w_j^3d_id_j/w_i}}{4w_iw_jd_id_j}
\eeq
For an undirected graph, $\frac{\partial \tildl_{ij}}{\partial w_i}|_{w_{1:n} = 1} = 0$ always. $\blacksquare$\\


\begin{prop}\cite{magnus1988matrix}\label{prop:12}
$\frac{\partial \lambda_k}{\partial L{ij}} = \sum_{i,j\neq t} v_iv_j$, where $v_i$ is the $i$-th element of the $k$-th eigenvector of $L$.
\end{prop}
\begin{prop}\label{prop:3}
Define $A$, $d$, $\lambda_{k}$, $L$, $\tildl$, $w$ as usual. Assume $\lambda_{k}$ is of multiplicity $1$. Define transition matrix $P = D^{-1}A$, $v_i$ is the $i$-th element of the $k$-th eigenvector of $L$.  \\
In symmetric perturbation, 
\beq
\frac{\partial \tilde{\lambda}_k}{\partial w_t}|_{w_{1:n} = 1} =  (1-\lambda_k)( \sum_i v_i^2 P_{it} - v^2_t)
\eeq
\end{prop}
{\bf Proof:}
After performing the symmetric perturbation, we obtain $\tildl$ from Equation \ref{eq:lhat1}.
We then calculate the influence function $\frac{\partial \tildl_{ij}}{\partial w_t}|_{w_{1:n} = 1}$,
\beq
\frac{\partial \tildl_{ij}}{\partial w_t}|_{w_{1:n} = 1} = \begin{cases}
									-\frac{A_{tj} + A_{jt}}{4\sqrt{d_td_j}} (1- \frac{A_{jt}}{d_j}) & i=t, j\neq t\\
									\frac{A_{ij} + A_{ji}}{4} \times \frac{A_{jt}d_i + A_{it}d_j}{(d_id_j)^{3/2}} & i, j \neq t\\
									0 & i, j = t\\
									\end{cases}
\eeq

Then we can derive $\frac{\partial \lambda_k}{\partial w_t}|_{w_{1:n} = 1}$ using Proposition \ref{prop:12},
 \begin{align}
 \frac{\partial \tilde{\lambda}_k}{\partial w_t}|_{w_{1:n} = 1} &=  \sum_{ij}\frac{\partial \tilde{\lambda}_k}{\partial \tildl_{ij}}\frac{\partial \tildl_{ij}}{\partial w_t} \\
&=\sum_{i, j \neq t} v_iv_j \frac{A_{ij} + A_{ji}}{2} \times \frac{A_{it}d_j + A_{jt}d_i}{2(d_id_j)^{3/2}} - v_t\sum_{j=1}^n v_j \times \frac{A_{tj} + A_{jt}}{2}(\frac{1}{\sqrt{d_jd_t}} - \frac{A_{jt}}{d_j\sqrt{d_jd_t}})\\
 & =  (1-\lambda_k)( \sum_i v_i^2 P_{it} - v^2_t),
\end{align}
$\blacksquare$\\

\subsection{Clustering}
Clustering, defined as a partition of nodes, cannot be written as a smooth function of weights, since the clustering can only be changed discretely. Therefore we cannot use IF to measure the robustness of clustering. However, a clustering can be evaluated by breakdown points (\BP), which will be discussed in the next section.\\
     
\section{Breakdown points (\BP)}
While IF measures local infinitesimal influences, the \mydef{breakdown point} (\BP) measures the global reliability of the graph properties. It is developed by \cite{Hampel:86} and widely used in robust statistics literature. It informs the range of perturbations that can be tolerated before the ``structural information in the data'' is lost. Similar with Influence function, the use of \BP~is limited by the assumption that the data is generated i.i.d from known distributions, thus not applicable to the graph properties. \\

We are the first to extend the definition of \BP~to the graph properties. For a graph property $f(\G)$, a \BP~is defined as,
\beq
 \sigma_w^* := max \{\sigma_w; |f(\Gw) - f(\G)| \le \epsilon \text{ with probability }1-\alpha\} 
\eeq
where $\epsilon$ and $\alpha$ are defined by users. Through finding \BP, we define meaningful and computable thresholds where information about a specific graph property is lost. For instance, for $f(\G,w)=\lambda_2(L(\G_w))$, a \BP~ can be defined when the eigengap between $\lambda_2$ and the next largest eigenvalue vanishes.\\

Another advantage of \BP~is that it allows robustness measure for non-differentiable graph property functions, such as clustering. It does not require the graph properties to be written as smooth functions of weights. It is a descriptive measure that allows global measure of robustness.



\section{Related Work} \label{sec:related}
In the past literature, there has been some work in perturbing the social networks. In \cite{karrerLN:08}, they restrict their perturbed networks to maintain the same number of vertices and edges as the original unperturbed network, and the perturbation is meant for the position of the edges only. The amount of perturbation is controlled by the number of edges being moved. In \cite{bhattacharyya2015subsampling} and  \cite{ali2016comparison}, they focus on subsampling the nodes of the networks. \cite{bhattacharyya2015subsampling} propose uniform subsampling bootstrap scheme, in which they iteratively select a subset of vertices without replacement and consider the graph induced by the subset of vertices. They also consider a subgraph subsampling bootstrap, where they use  a enumeration scheme to find all possible subgraphs with a fixed vertice size, and they selecte the subgraph with a fixed probability $p$. \\

We have also seen work in detecting dense communities and largest connected component in \cite{verzelen2015community}. They formalize tests for the existence of a dense random subgraph based on a variant of scan statistics. Although they offer sharpe detection bounds, their theorems only make a judgement on the existence of the subgraph instead of finding out where the subgraph is. Moreover, one has to go through all the subgraphs in order to calculate the test statistics, which in application, could be computationally intensive.

\section{Experiments} \label{sec:5}
In this section, we perform experiments to test the robustness of clustering, WCut, number of wCC's and eigengap. We employ both synthetic datasets and a Facebook dataset. Because symmetric perturbation is less interpretable, we only apply asymmetric perturbation when studying connected components and eigengap, and apply asymmetric perturbation for the remainder of the study.
\subsection{Datasets}
\begin{figure}[ht]
\begin{tabular}[b]{ccc}
\includegraphics[width=0.5\linewidth]{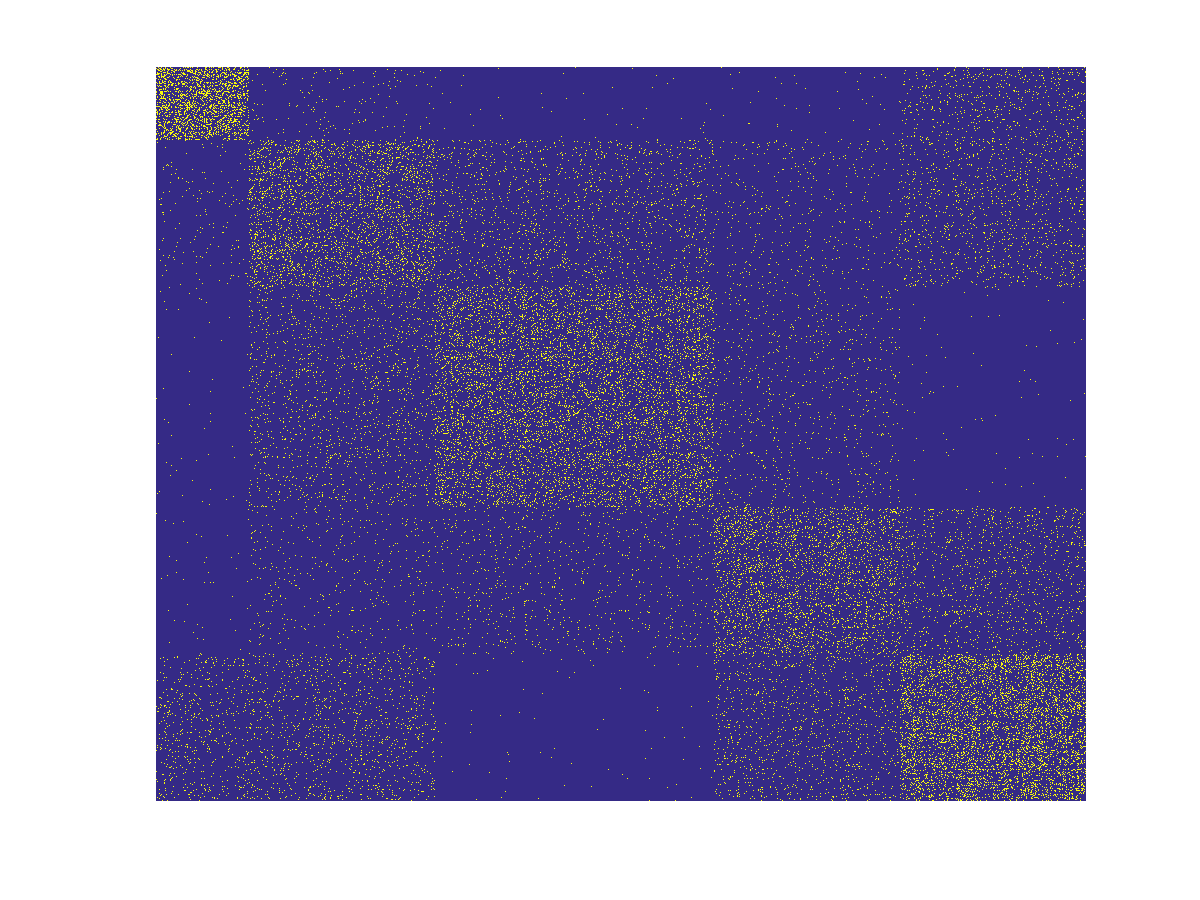} &
\includegraphics[width=0.5\linewidth]{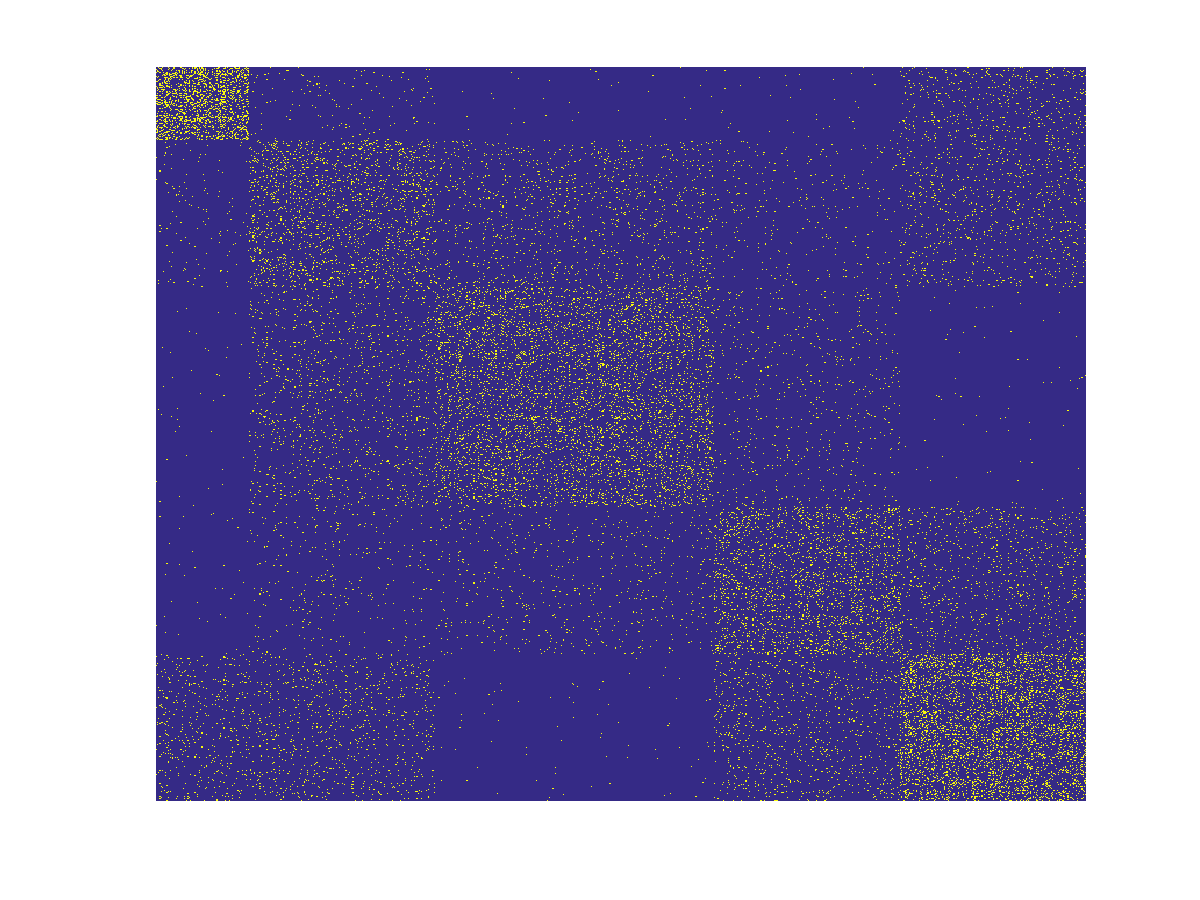} 
\\
\includegraphics[width=0.5\linewidth]{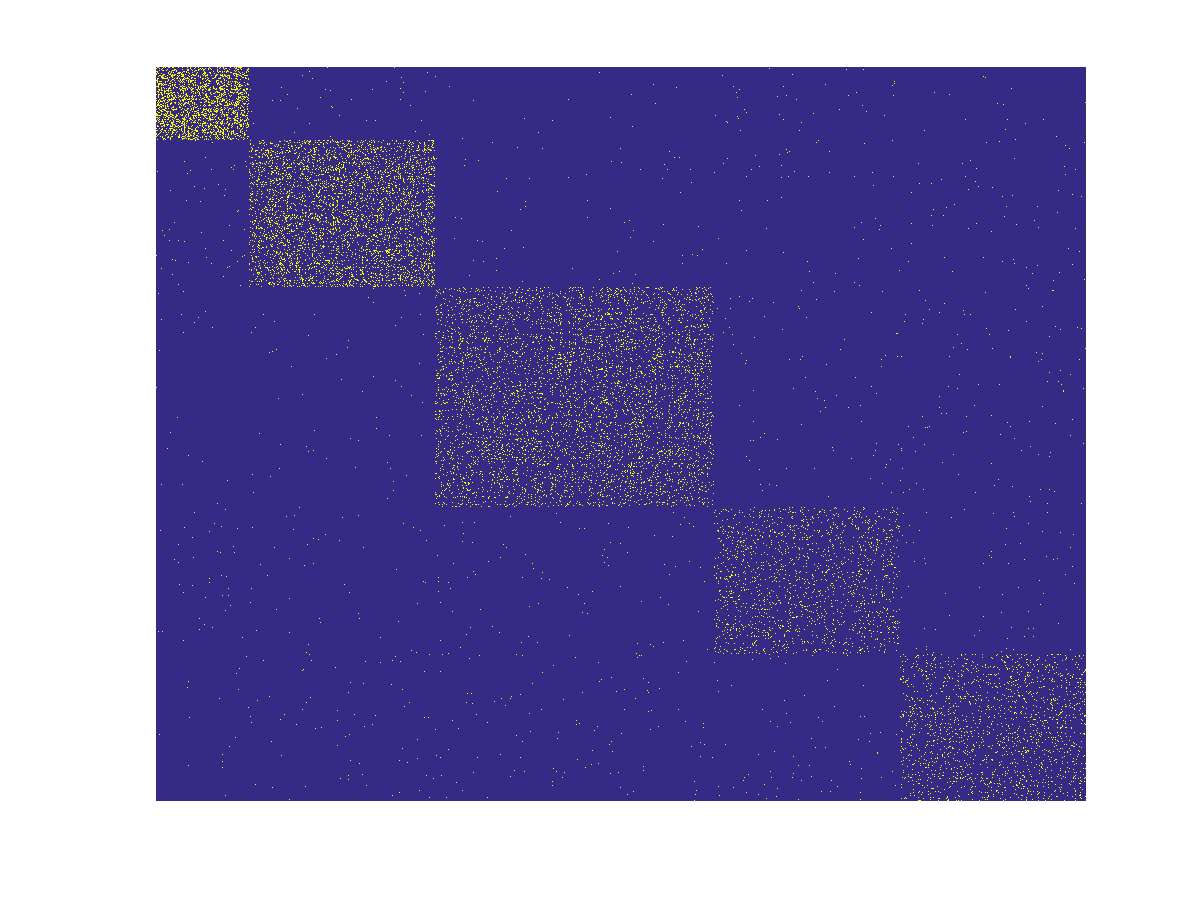} &
\includegraphics[width=0.5\linewidth]{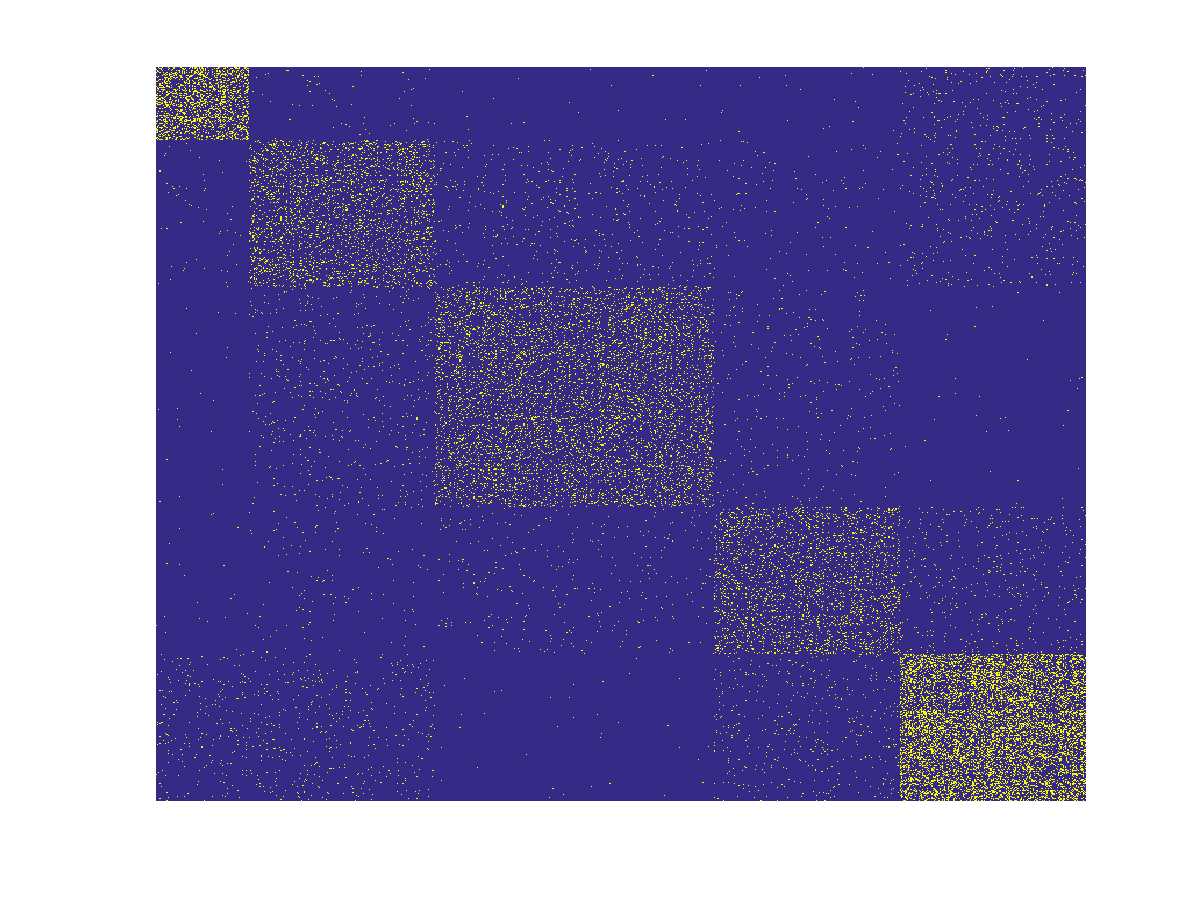}
\end{tabular}
\caption{\label{fig:1syn}The synthetic datasets. Top left: $n=800$,  $\lambda_{1:K} = (0, 0.2, 0.4, 0.6, 0.8)$, $w_{DC-SBM}^i \sim 0.5 + 0.5\times Uniform(0, 1)$. Top right: $n$ and $\lambda_{1:K}$ the same with top left, $w_{DC-SBM}^i \sim 0.4 + 0.6\times Uniform(0, 1)$. Down left: $n=2000$, $\lambda_{1:K} = (0, 0.1, 0.11, 0.12, 0.13)$. Down right: $n=800$, $\lambda_{1:K} = (0, 0.1, 0.2, 0.3, 0.4)$}
\end{figure}

{\bf Synthetic datasets} The synthetic datasets are generated from the DC-SBM model with the number of clusters $K = 5$, and the distribution of the cluster sizes $\frac{n_k}{n} = (0.1, 0.2, 0.3, 0.2, 0.2)$. DC-SBM is defined as a class of network model that $A_{ij}=w_iw_jB_{kl}$ whenever $i\in k,j\in l$, with $B=[B_{kl}]\in\rrr^{K\times K}$ symmetric and non-negative and $w_1,\ldots, w_n$ non-negative weights associated with the graph nodes. . The clustering is guaranteed to be recovered with small errors by spectral clustering algorithm if the graph is generated from DC-SBM \cite{qin2013regularized, WanM:isaim16}. The weights of DC-SBM, $w^i_{DC-SBM} \sim 0.5 + 0.5\times Uniform(0, 1)$ if not otherwise specified. The graphs are generated with different $n$ and spectrum in the following experiments. The visualization of the graphs and the model parametrization are shown in Figure \ref{fig:1syn}.\\

\begin{figure}[ht]
\includegraphics[width=0.8\linewidth]{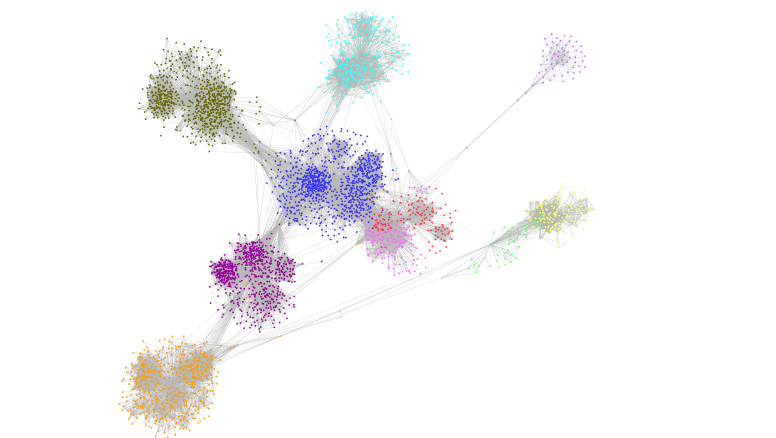}
\caption{\label{fig:7} Facebook dataset}
\end{figure}

{\bf Facebook dataset } The Facebook dataset \cite{leskovec2012learning} is an undirected connected graph which consists of 10 anonymized ego networks. It has 4039 nodes and 88234 edges. The data was collected from survey participants using a Facebook application called Social Circles. Each cluster is consists of the members within an ego network. The visualization of the Facebook dataset is shown in Figure \ref{fig:7}.\\
In our experiments, we only examines undirected graphs. Since the definition of $L$ is universal for all graphs, the robustness of WCut and clustering can be measured in the same way. The graph properties we defined for the number of wCC's and eigengap are only meaningful for undirected graphs, and we do not know yet how to measure their robustness for directed graphs.\\

We say a dataset or a graph is hard if there are many edges between clusters, and graph properties calculated from them are expected to be sensitive. A harder graph usually has smaller $n$, larger eigengap, denser connections between clusters and sparser connections within clusters. From the theory of the recovery of clustering \cite{MWan:nips15, WanM:isaim16, WanM:nips16}, it can be indicated that the graph properties in the harder graphs will be less robustness. \\

Notice that in the real datasets, e.g. Facebook, one may argue that the perturbation may not be meaningful since the edges are given continuous weights after the perturbation, while the edges can only take the values $0$ and $1$, e.g, two people are either friends or not. It is true that in reality, the change of this kind of social networks is restricted to adding or deleting nodes or edges. Our perturbation methods do allow for deletion of nodes and edges when the weight are generated node resampling mechanism. They can be viewed as deletions from the graph. We can also utilize node resampling to generate discrete weights for the nodes and edges to make it more meaningful. One potentail concern is that we do have the constraint that no new edges or nodes can be added from our perturbation framework. \\

\subsection{Robustness of clustering}
We design this experiment to answer two questions. Firstly, For fixed $\sigma_w$, does \BP~depend on weight distribution? Secondly, Is \BP~informative? We perform the experiment using the following steps. Firstly, we generate $w$ from the four weight distributions: node resampling, binary, gamma and mixture distributions. with $E(w) = 1$ and $\sigma_w$, as described in Section \ref{sec:2}.
We assign various magnitude of $\sigma_w$ to capture the amount of perturbation the clusterings can tolerate. \\

Secondly, for each perturbed graph $\Gw$, we perform spectral clustering algorithm and obtain $\tilde{\clust}$. We then compute misclassification errors $\distance(\clust_{true}, \tilde{\clust})$, and $\distance(\clust_{spc}, \tilde{\clust})$, where $\clust_{true}$ is the true clustering of the underlying model, and $\clust_{spc}$ is the clustering obtained from the unperturbed graph through spectral clustering. The misclassification error is define in Section \ref{sec:def1}. If there is a value of $\sigma_w$ where $\tilde{\clust}$ becomes very unstable, that the misclassification error starts to have high variance, we will call it the breakdown point. \\

\begin{figure}[ht]
\begin{tabular}[b]{cc}
\centering
\raisebox{-0.5\height}{\includegraphics[width=0.35\linewidth]{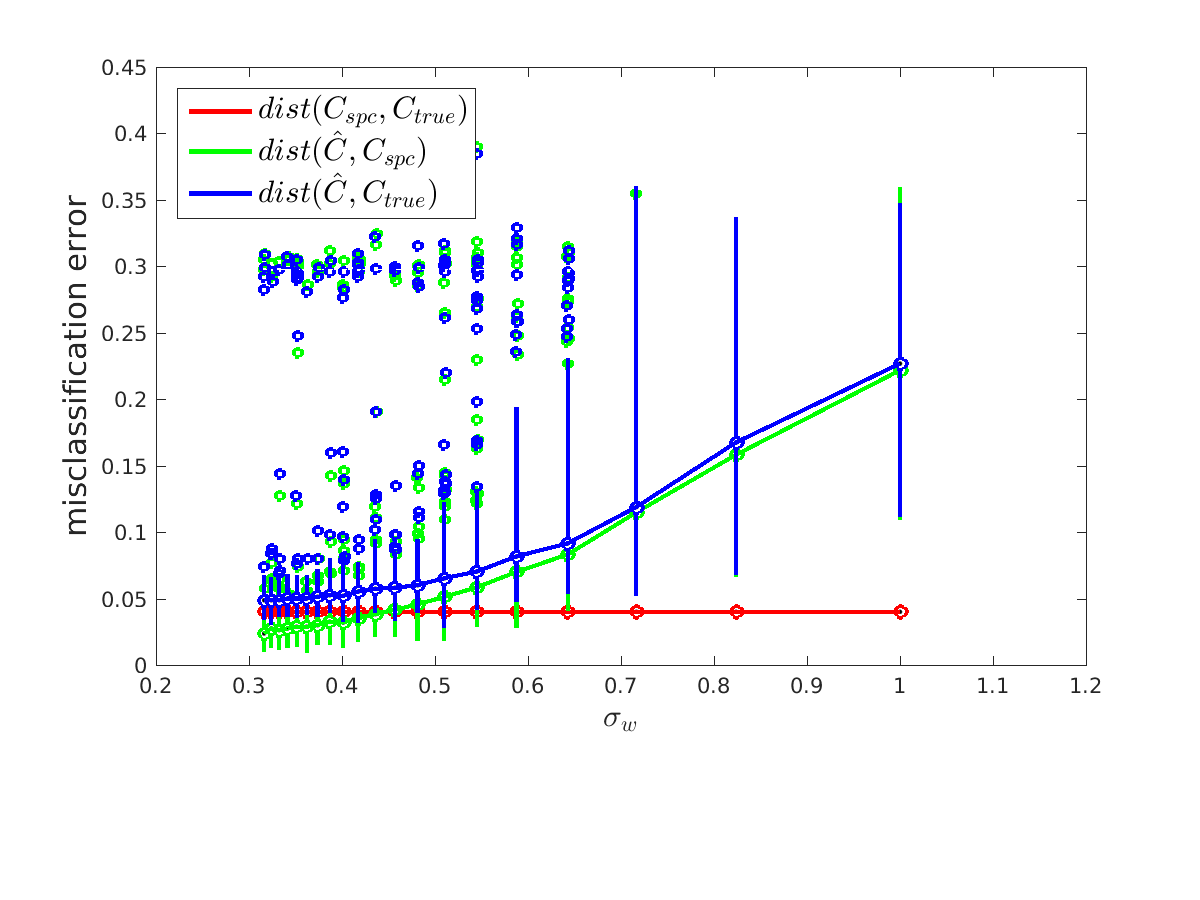}}&
\raisebox{-0.5\height}{\includegraphics[width=0.35\linewidth]{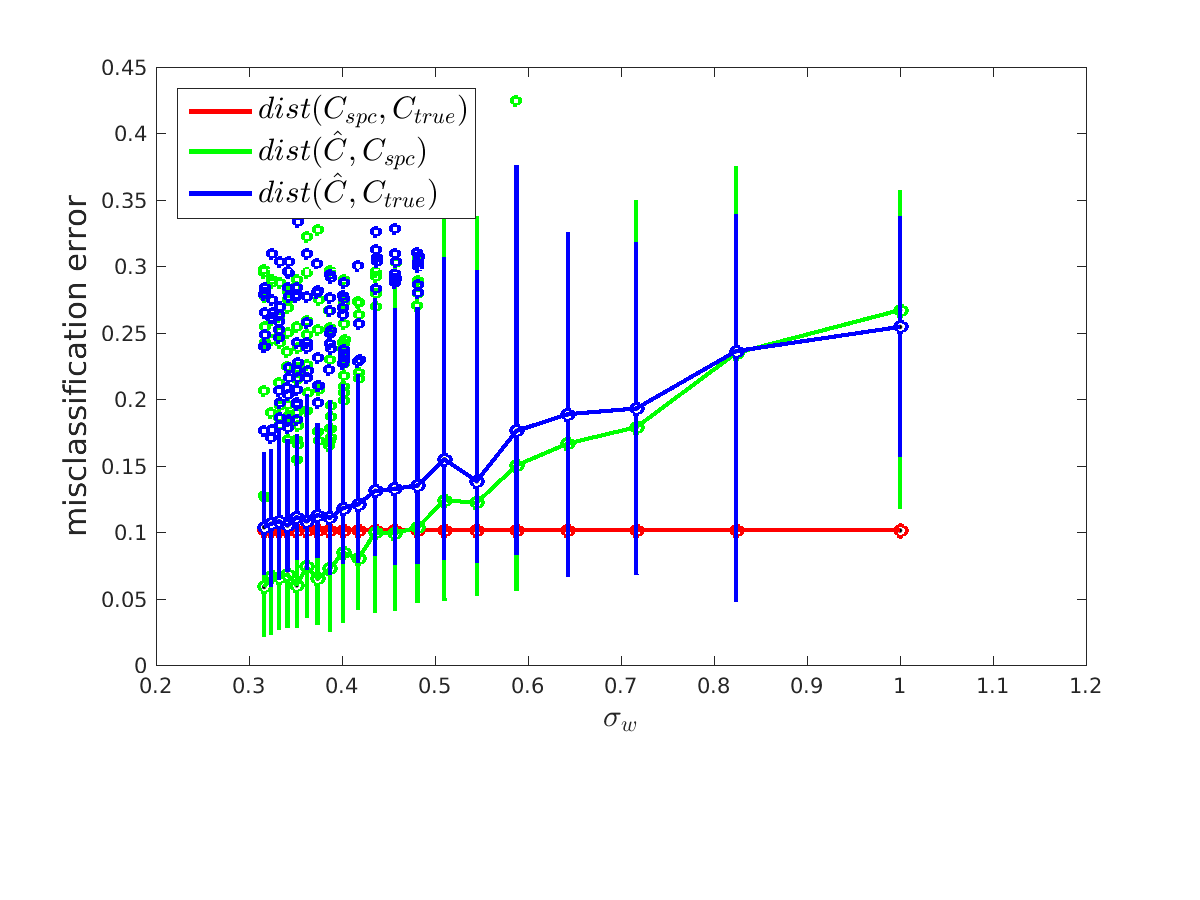}}\\
\raisebox{-0.5\height}{\includegraphics[width=0.35\linewidth]{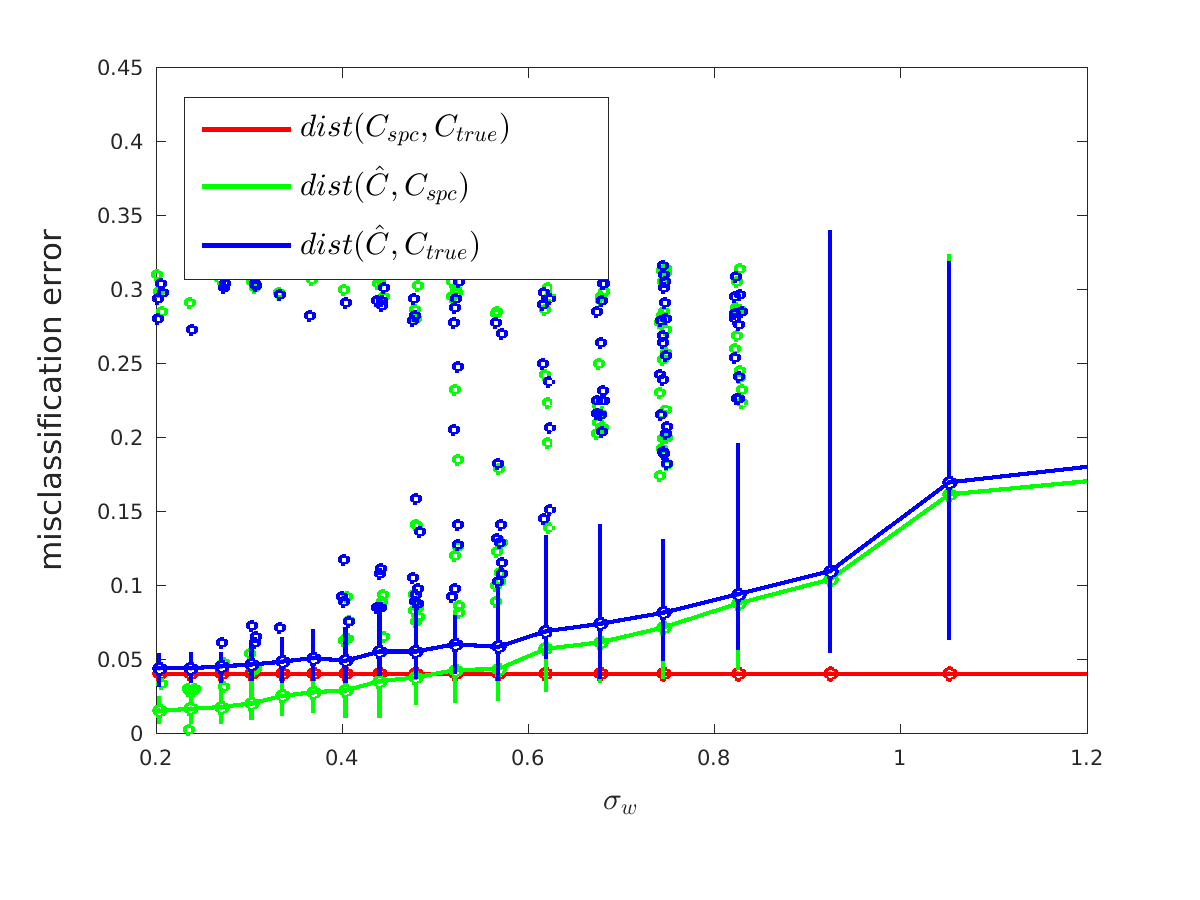}}&
\raisebox{-0.5\height}{\includegraphics[width=0.35\linewidth]{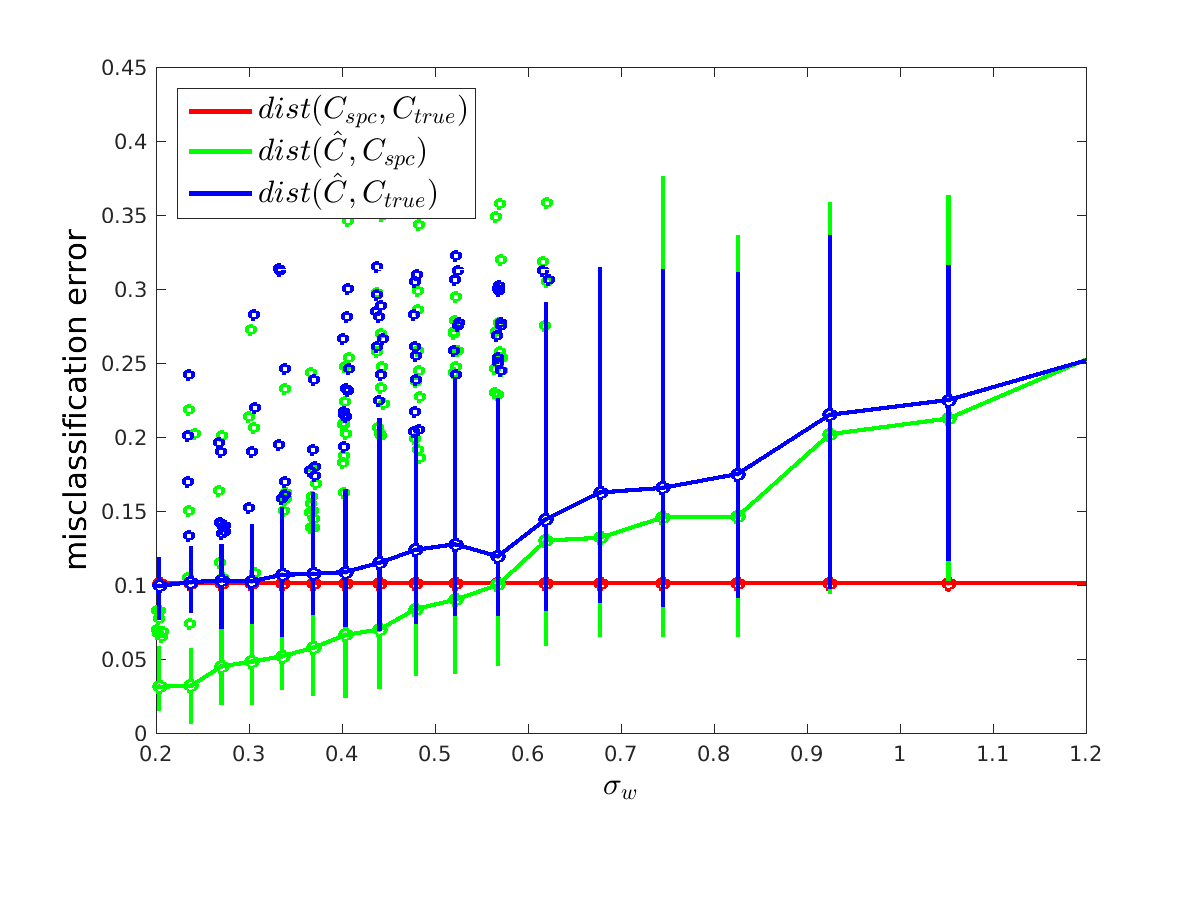}}\\
\raisebox{-0.5\height}{\includegraphics[width=0.35\linewidth]{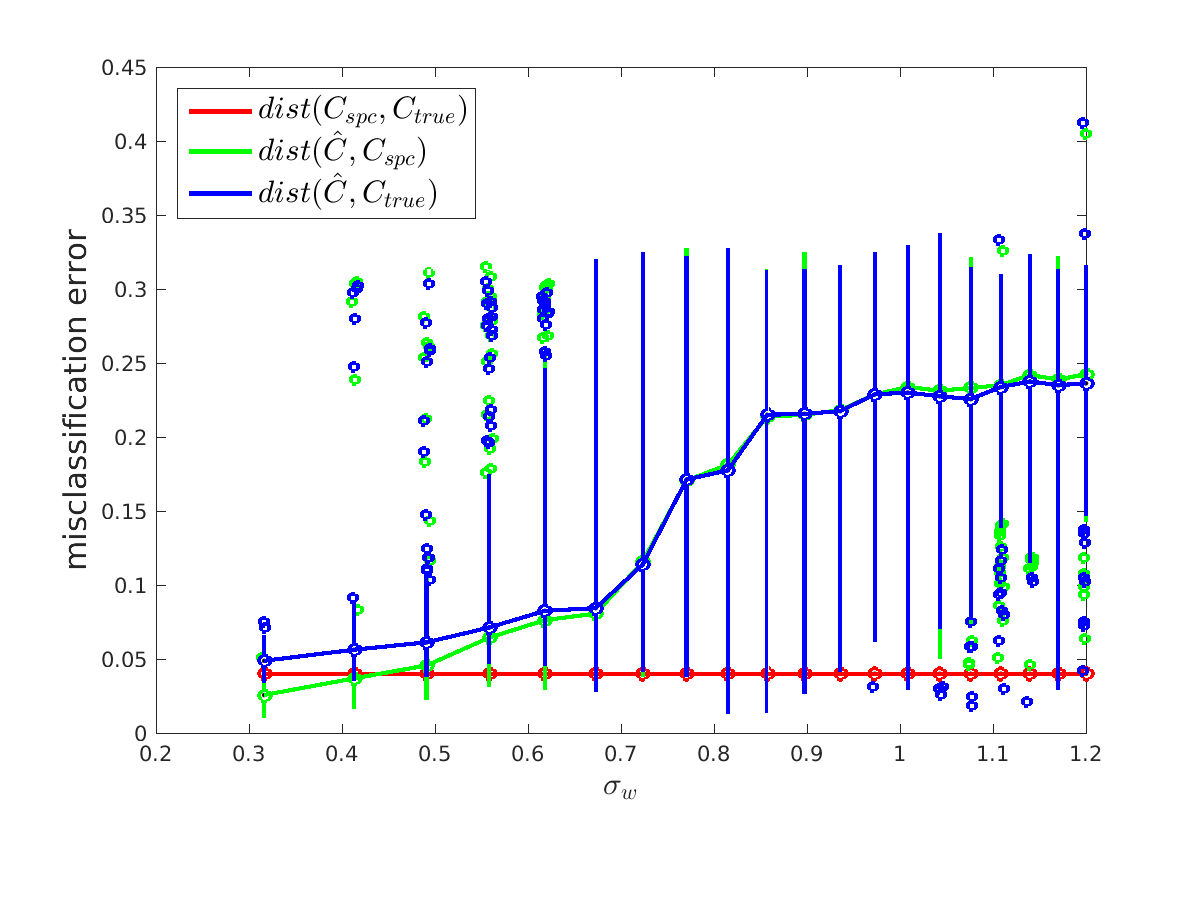}}&
\raisebox{-0.5\height}{\includegraphics[width=0.35\linewidth]{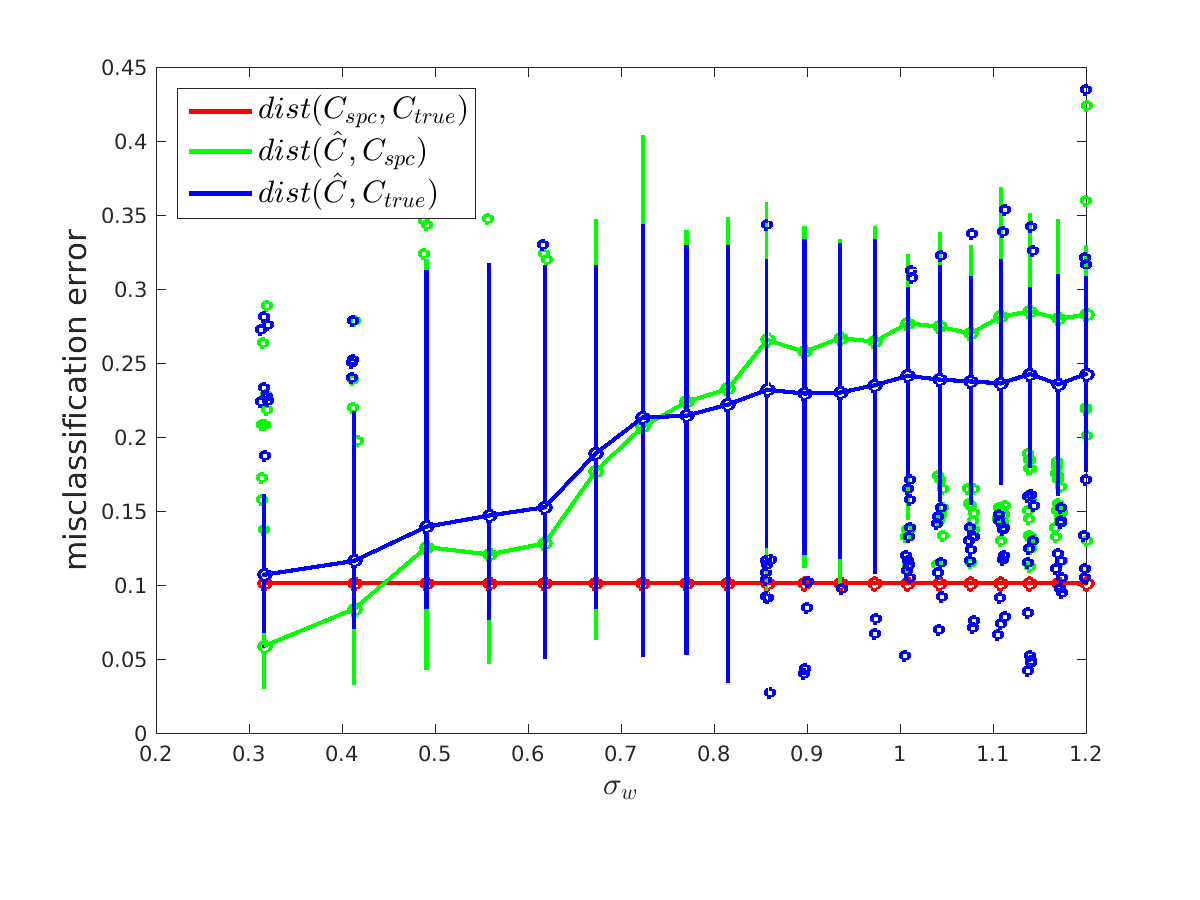}}\\
\raisebox{-0.5\height}{\includegraphics[width=0.35\linewidth]{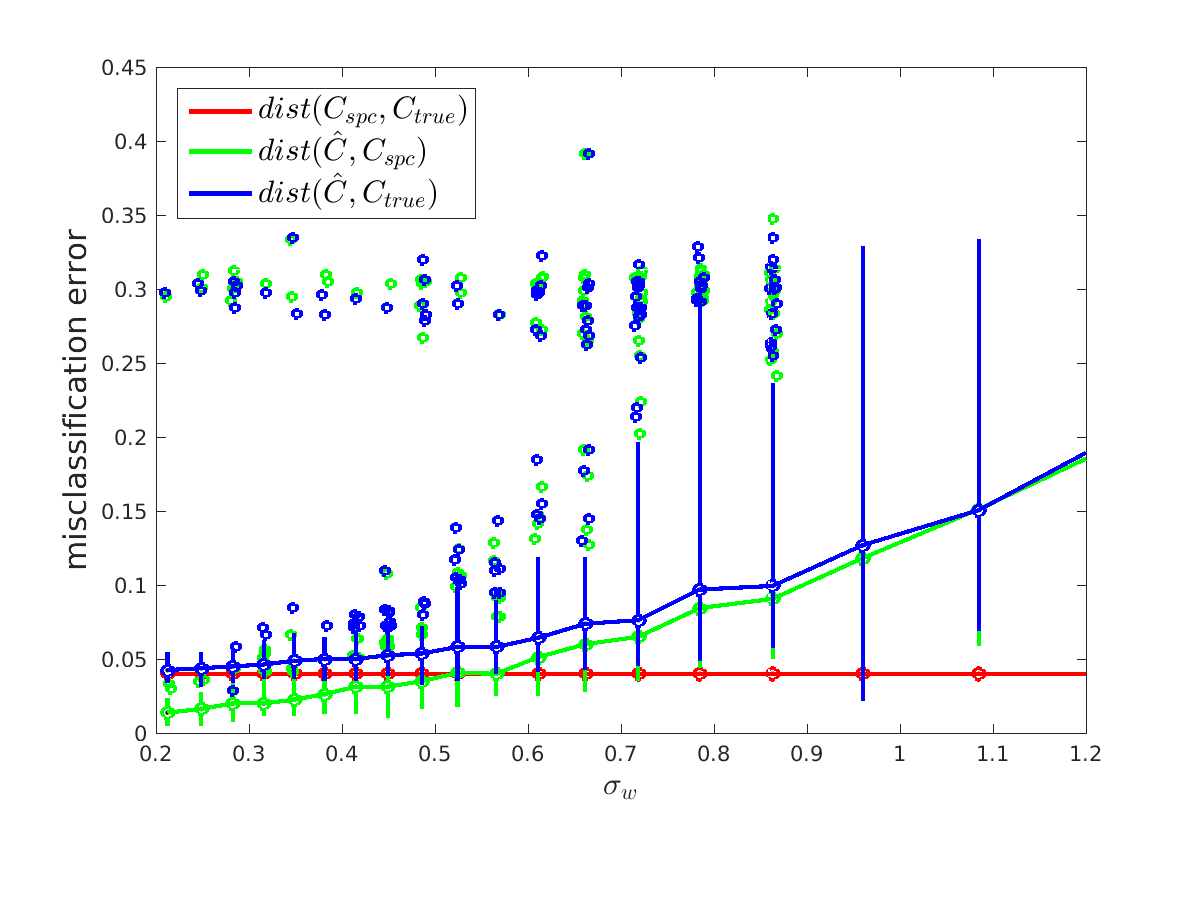}}&
\raisebox{-0.5\height}{\includegraphics[width=0.35\linewidth]{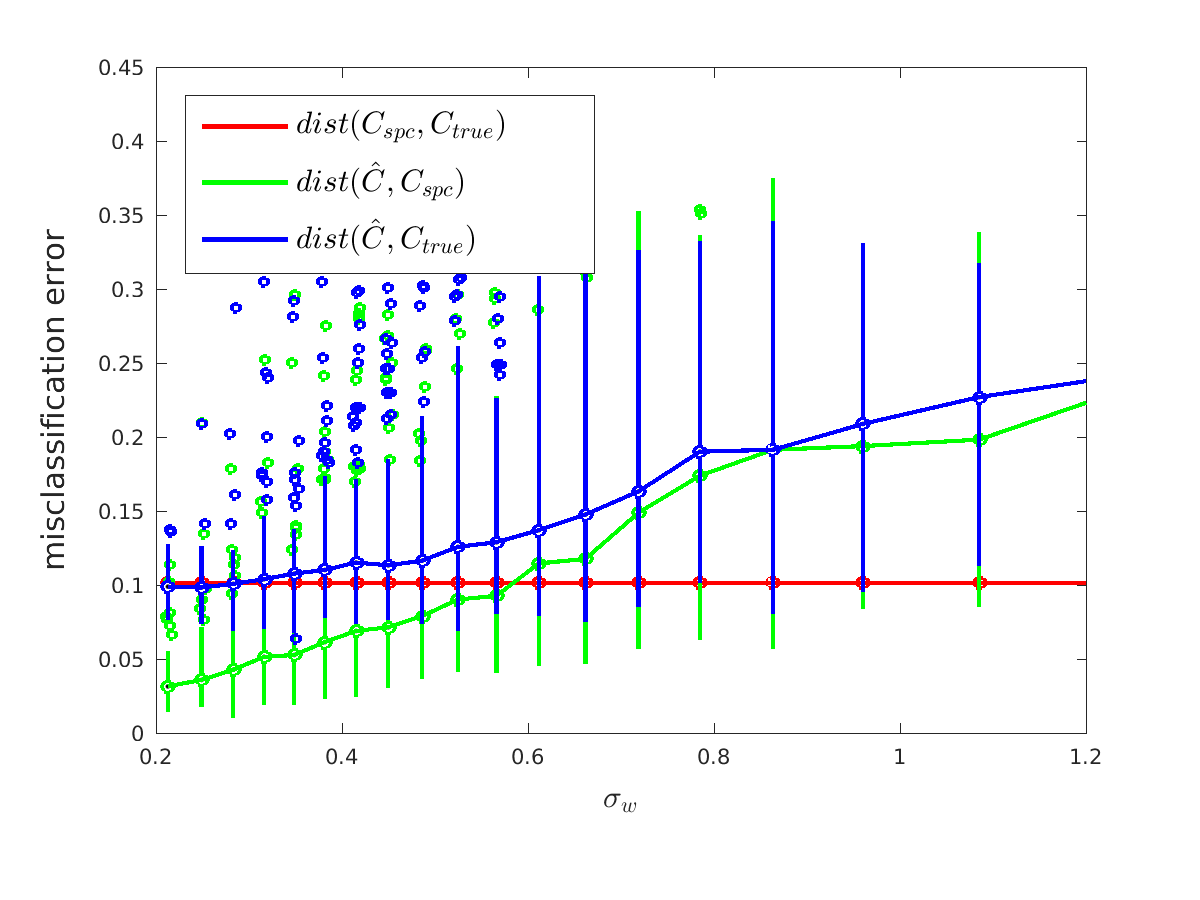}}\\
\end{tabular}
\caption{\label{fig:2} Left: easy dataset with $w_{DC-SBM}^i \sim 0.5 + 0.5\times Uniform(0, 1)$. Right: hard dataset with $w_{DC-SBM}^i \sim 0.4 + 0.6\times Uniform(0, 1)$. $1$st row: Node resampling. $2$And row: Binary . $3$rd row: gamma distribution. $4$th row: mixture-uniform gamma distirbution. $\tilde{\clust}$ is the weighted version of clustering obtained from spectral clustering algorithm. Each boxplot is consist of 100 repetitions.}
\end{figure}

We employ two undirected synthetic datasets from DC-SBM. They both have $n=800$, and $\lambda_{1:K} = (0, 0.2, 0.4, 0.6, 0.8)$. The difference is that the first dataset is generated with $w_{DC-SBM}^i \sim 0.5 + 0.5\times Uniform(0, 1)$, and the second one is generated with $w_{DC-SBM}^i \sim 0.4 + 0.6\times Uniform(0, 1)$. The results are shown in Figure \ref{fig:2}.\\

For all the cases, we observe a significant break in the variance of misclassification errors, which we define as \BP. For example, the \BP~for binary distribution in the easy dataset is around $0.8$, which becomes $0.6$ in the hard dataset. Notice that with all weight distributions, \BP's in the harder dataset are always smaller comparing to that in the eaiser dataset. This confirms that \BP~is informative, since it predicts the sensitivity of graph properties in harder (less robust) dataset. For the same dataset, across different weight distributions, the misclassification errors break at different $\sigma_w$, which suggests that \BP~varies with different weight distributions. 

\subsection{Robustness of WCut}
In the experiments here we want to verify that IF of WCut is informative, and then probe the source of robustness of WCut at the node level.\\

\begin{figure}
\begin{tabular}[b]{cc}
\includegraphics[width=0.35\linewidth]{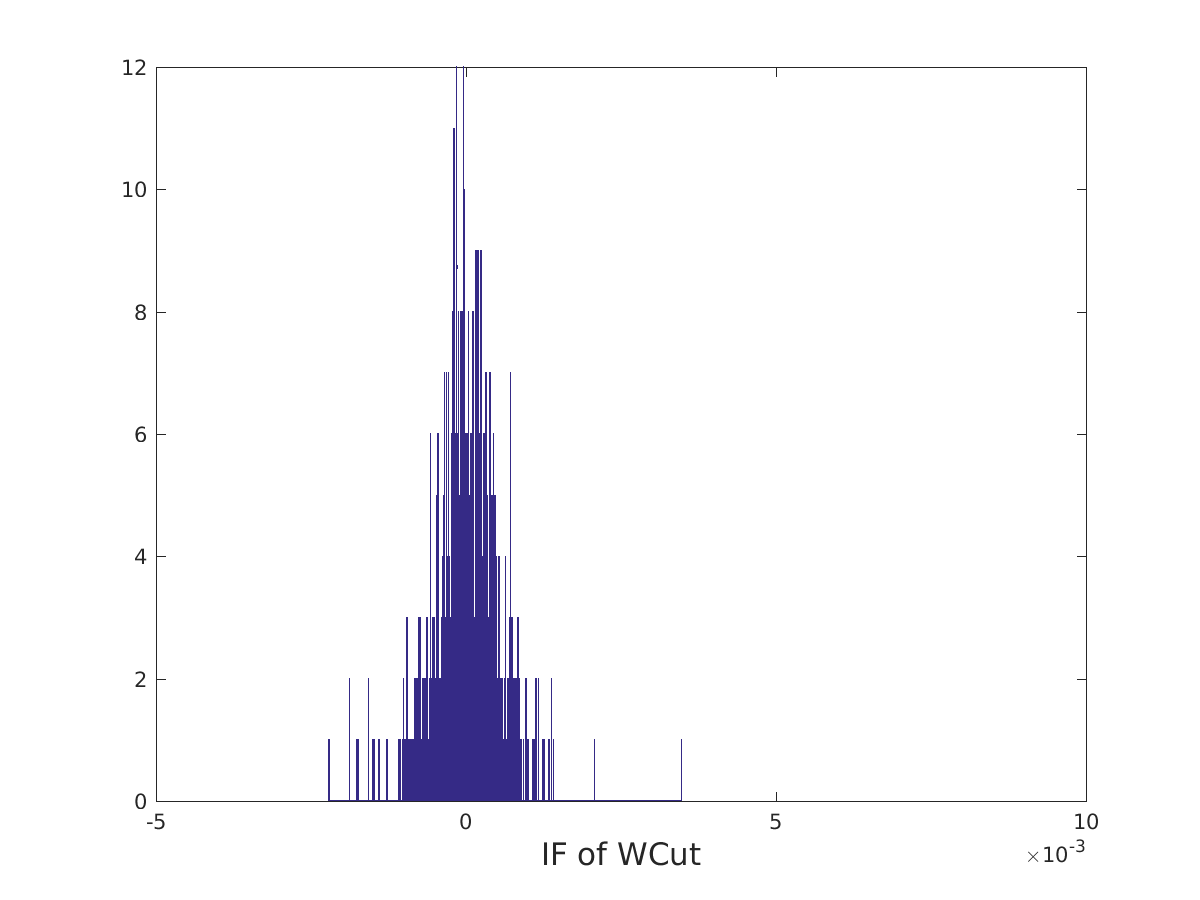} &
\includegraphics[width=0.35\linewidth]{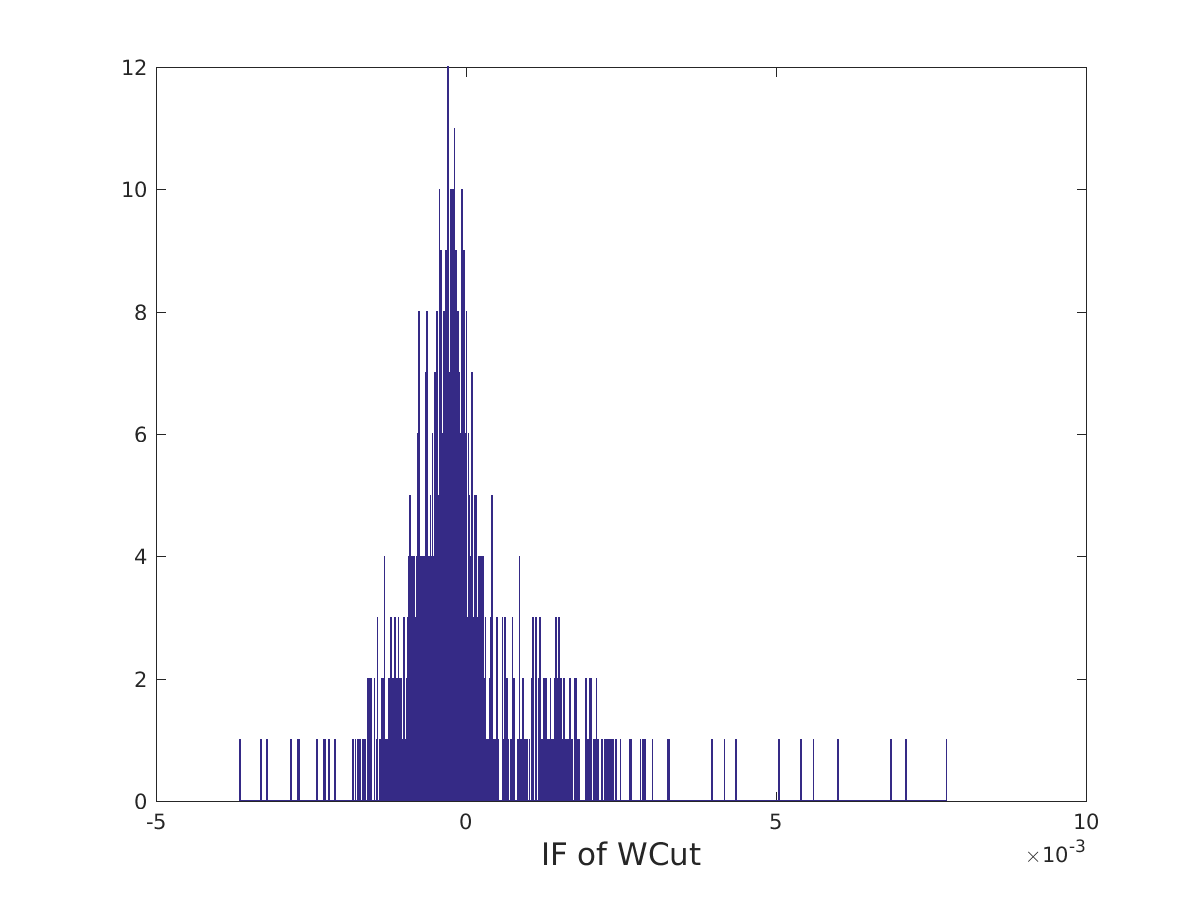}
\end{tabular}
\caption{\label{fig:hist} The histograms of $IF^{WCut}$ for synthetic datasets. Left: $\clust$. Right: $\tilde{\clust}$.}
\end{figure}

For the synthetic dataset with $n=800$, $w\sim 0.5+0.5\times Uniform(0, 1)$, We firstly obtain its clustering $\clust$ from spectral clustering algorithm. We then add noise to $\clust$ by randomly picking $200$ nodes and reassigning them to other clusters randomly, in which way we obtain $\tilde{\clust}$. The WCut of $\tilde{\clust}$ is expected to be less robust comparing to that of $\clust$, since there are already more noises in $\tilde{\clust}$. After computing $IF^{WCut}$ for both $\clust$ and $\tilde{\clust}$, we plot the histograms in Figure \ref{fig:hist}. We observe that in the histogram for $\clust$, $IF^{WCut}$ is more concentrated on $0$ with fewer large influence nodes, thus indicating a robust clustering. On the other hand, the histogram for $\tilde{\clust}$ indicates the clustering being sensitive. These findings correspond with our expectations and show that $IF^{WCut}$ is informative.\\ 

The other experiment we do is to probe where the sensitivity comes from through a partial perturbation. We generate synthetic dataset with $n=800$, and $\lambda_{1:K} = (0, 0.2, 0.4, 0.6, 0.8)$.  We also perform the experiment on the Facebook dataset. We select the nodes with $IF^{WCut}_i(\clust_{true}) > 0 $. These nodes are considered not well clustered because increasing the weights on them is expected to increase WCut, thus worse quality of clustering. For these nodes, we generate $w$ from $Gamma(0.1/\mu, 10\mu^2)$, where $E(w) = \mu$, $Var(w) = 0.1$. The rest of the nodes have $w=1$. We then assign the weights to the edges through asymmetric perturbation. We also examine the perturbed clustering from spectral clustering algorithm.\\

\begin{figure}[ht]
\begin{tabular}[b]{cc}
\includegraphics[width=0.5\linewidth]{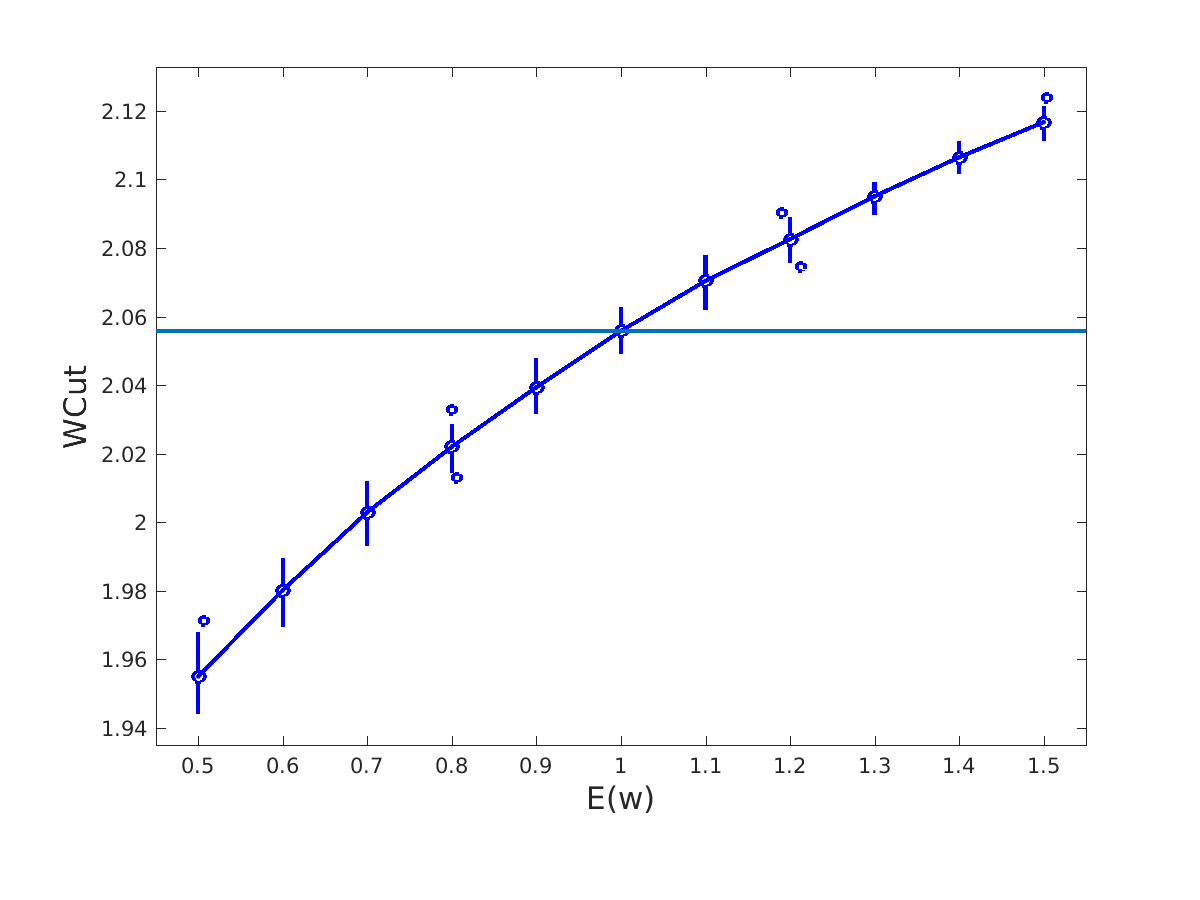}
\includegraphics[width=0.5\linewidth]{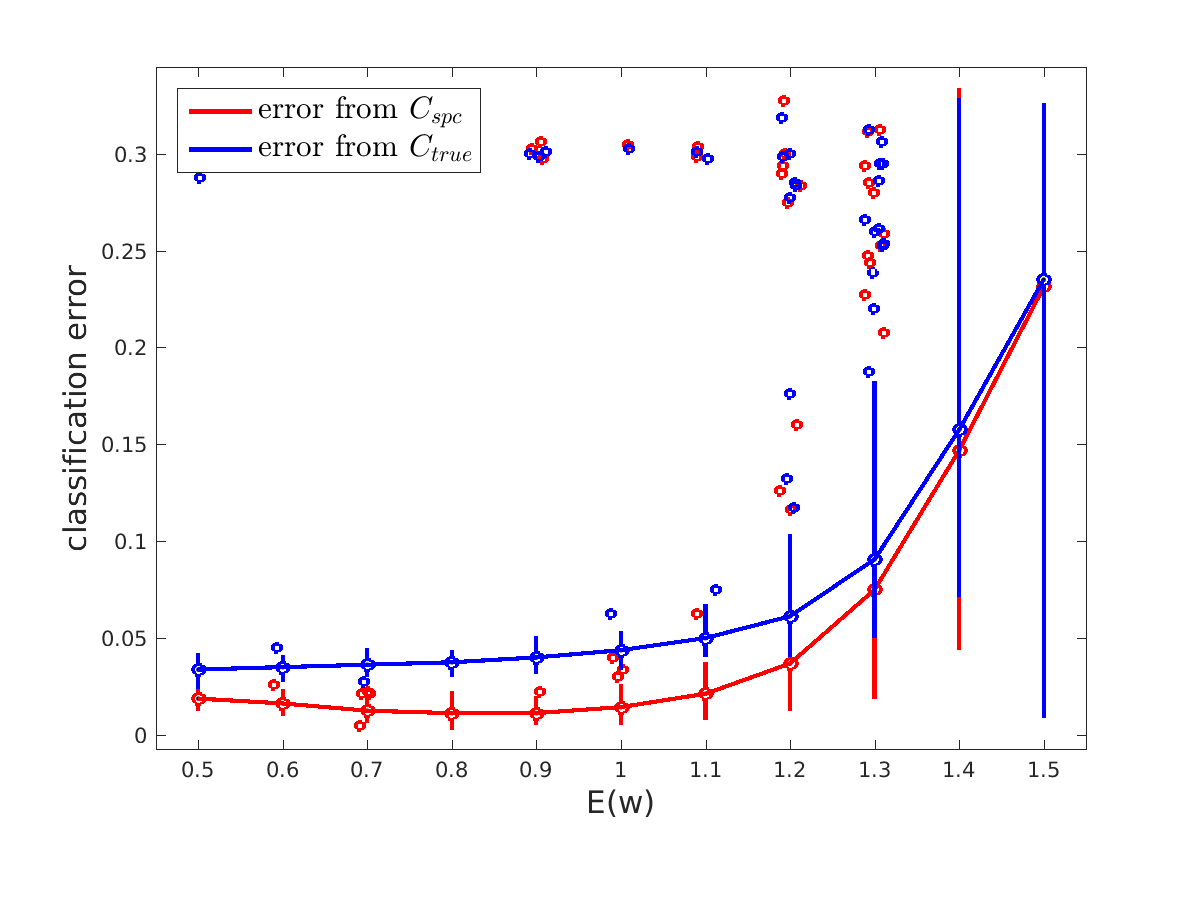}\\
\includegraphics[width=0.5\linewidth]{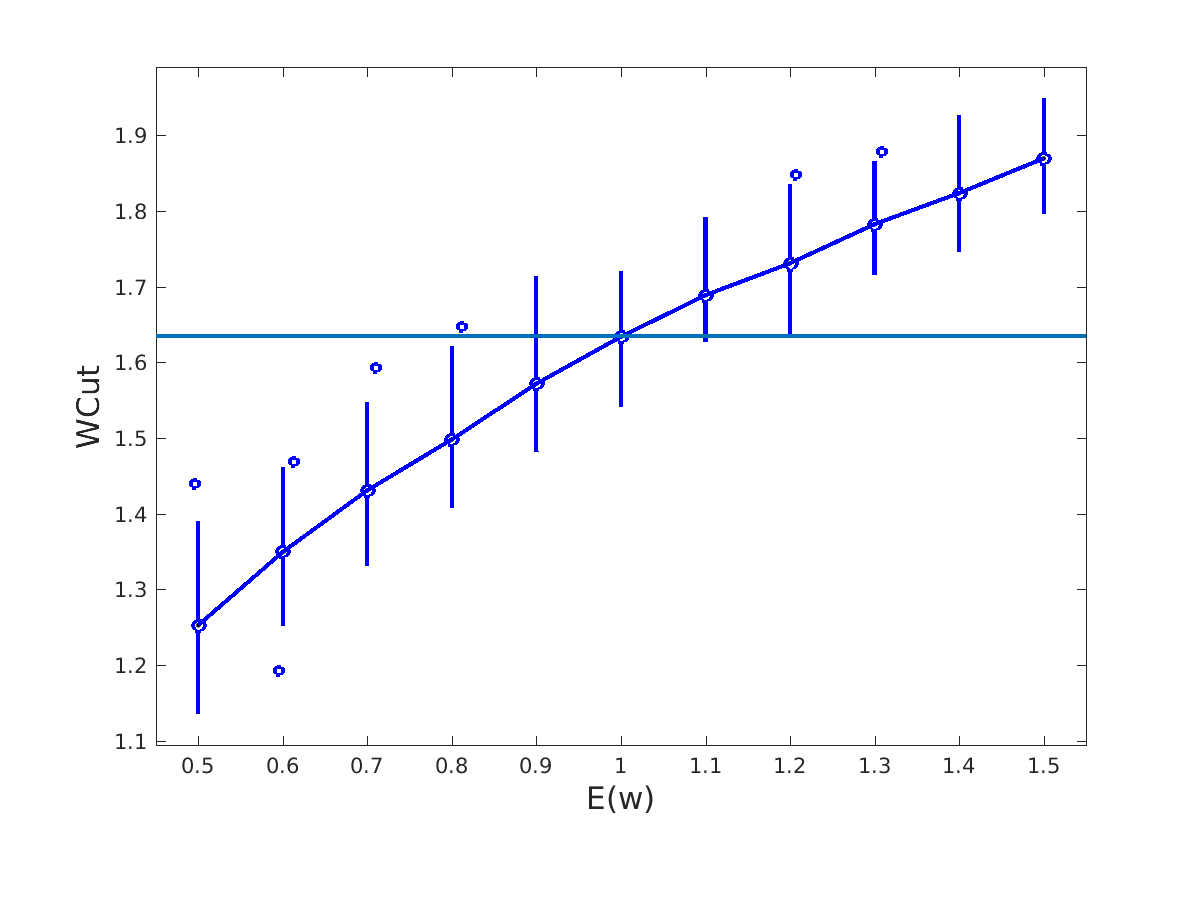}
\includegraphics[width=0.5\linewidth]{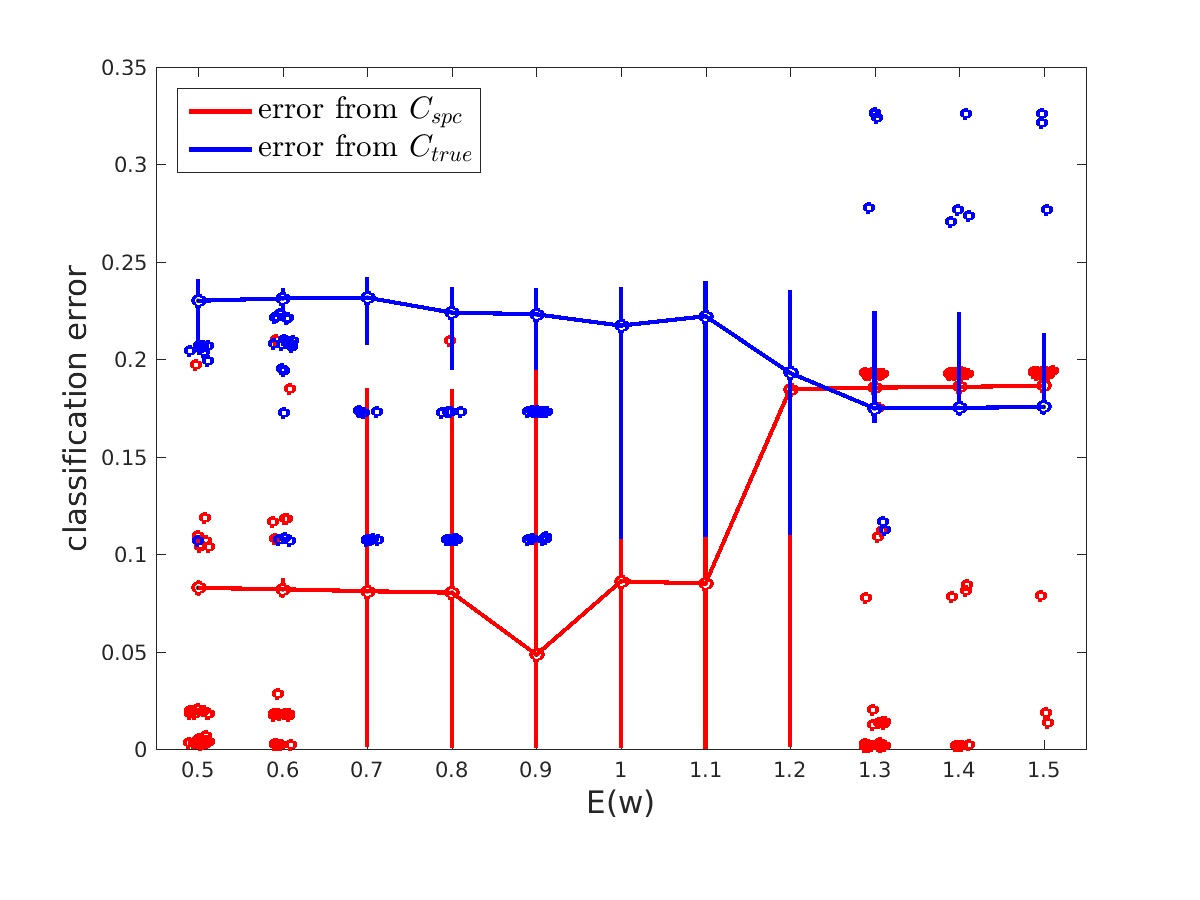}
\end{tabular}
\caption{\label{fig:4}Left column: the change of WCut with respect to $E(w)$. Right column: the change of classification error with respect of $E(w)$. First row: synthetic datset. Second row: Facebook datset. Each boxplot is consist of 100 repetitions.}
\end{figure}

The results are shown in Figure \ref{fig:4}, we observe that with both datasets, WCut increases as the weights on the nodes with bad influence increase, which indicates that the nodes with $IF^{WCut}$ causes the quality of clustering to drop, and could potentially be not well clustered. In the synthetic dataset, $\distance(\tilde{\clust}, \clust_{true})$ increases as $E(w)$ increases, and decreases slightly but steadily as $E(w)$ decreases. This is because spectral clustering is equivalent to optimizing WCut \cite{MPentney:sdm07},  when less weight is imposed on nodes with bad influence for $IF^{WCut}_i(\clust_{true})$, WCut gets smaller and $\tilde{\clust}$ becomes closer to $\clust_{true}$. In the Facebook dataset, $\distance(\tilde{\clust}, \clust_{true})$ decreases as $E(w)$ increases. This is because the underlying clustering $\clust_{true}$ in Facebook does not correspond to the clustering that minimizing WCut. Therefore minimizing WCut does not lead to $\tilde{\clust}$ being close to $\clust_{true}$.

\subsection{Robustness of wCC's and eigengap} \label{robustcc}
We firstly probe the sensitivity of wCC's and eigengap. We calculate $IF^{f_u}_i$, $IF^{f_l}_i$ and $IF^{f_e}$ for the nodes and select the nodes with bad influence. We call the nodes with $IF^{f_u}_i < 0$ or $IF^{f_l}_i > 0$ or $IF^{f_e} <0$ nodes with bad influence, since these nodes make the distinction between wCC's or eigengap less significant. We do a partial perturbation on the graphs by perturbing the weights of these nodes with $w_i \sim Uniform(a, a+0.25)$, $a \in [0.5, 0.6, \cdots, 1.5]$. \\

We also examine their robustness by making full perturbation on the entire graph. We generate $w_i \sim Mixture(0.5, b, 0, 0.1, T_-, Gamma, p)$ for all the nodes, where $T_-$ is a distribution centered around $0.51$ with probability $1$, $p$ ranges from $0.1$ to $0.9$. We choose $T_-$ to maintain the weights on the edges to be positive. Since the strength of perturbation is mostly coming from $T_+$ and the choice of the base binary distribution, we do not lose generosity. In this way $E(w) = 1$ and $\sigma(w) = \sigma_w$ varies.\\

\begin{figure}[ht]
\begin{tabular}[b]{ccc}
\includegraphics[width=0.5\linewidth]{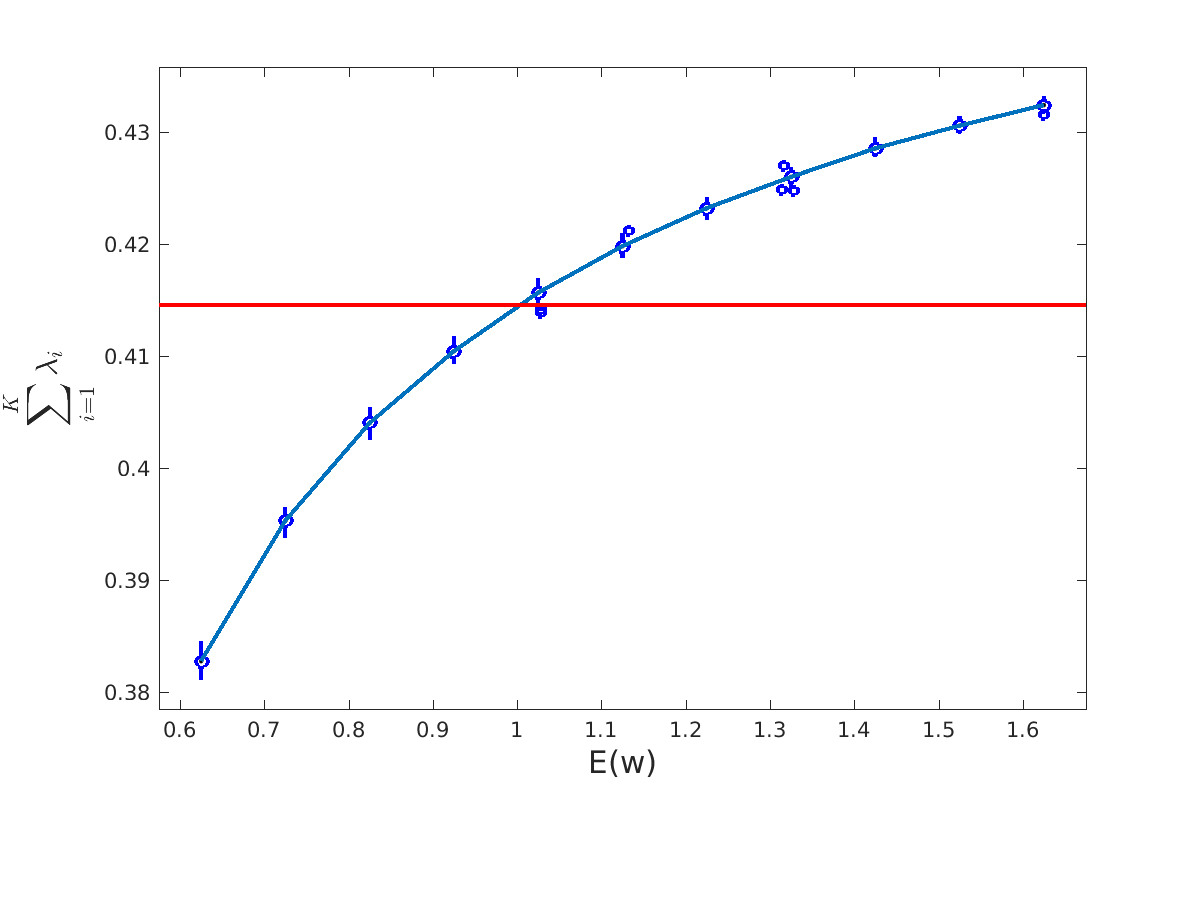}
\includegraphics[width=0.5\linewidth]{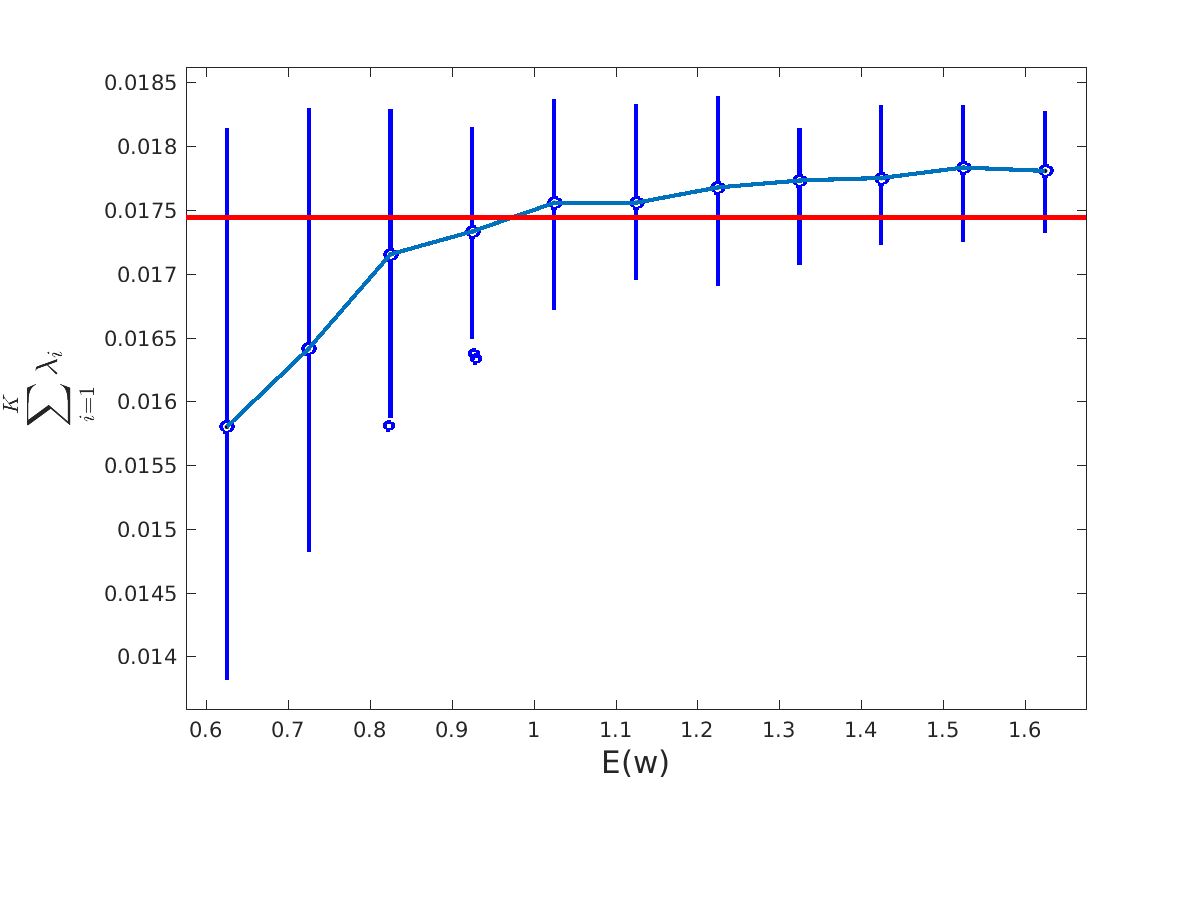}\\
\includegraphics[width=0.5\linewidth]{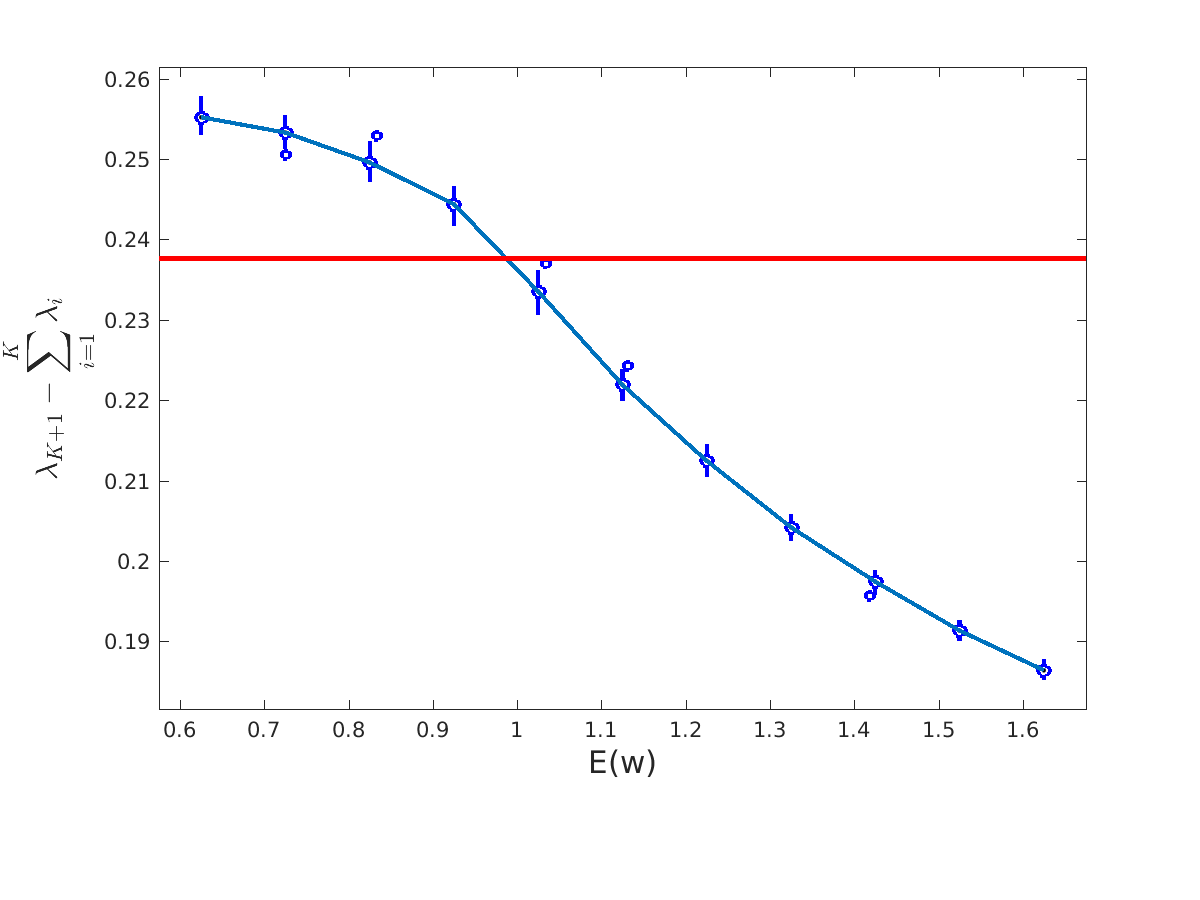}
\includegraphics[width=0.5\linewidth]{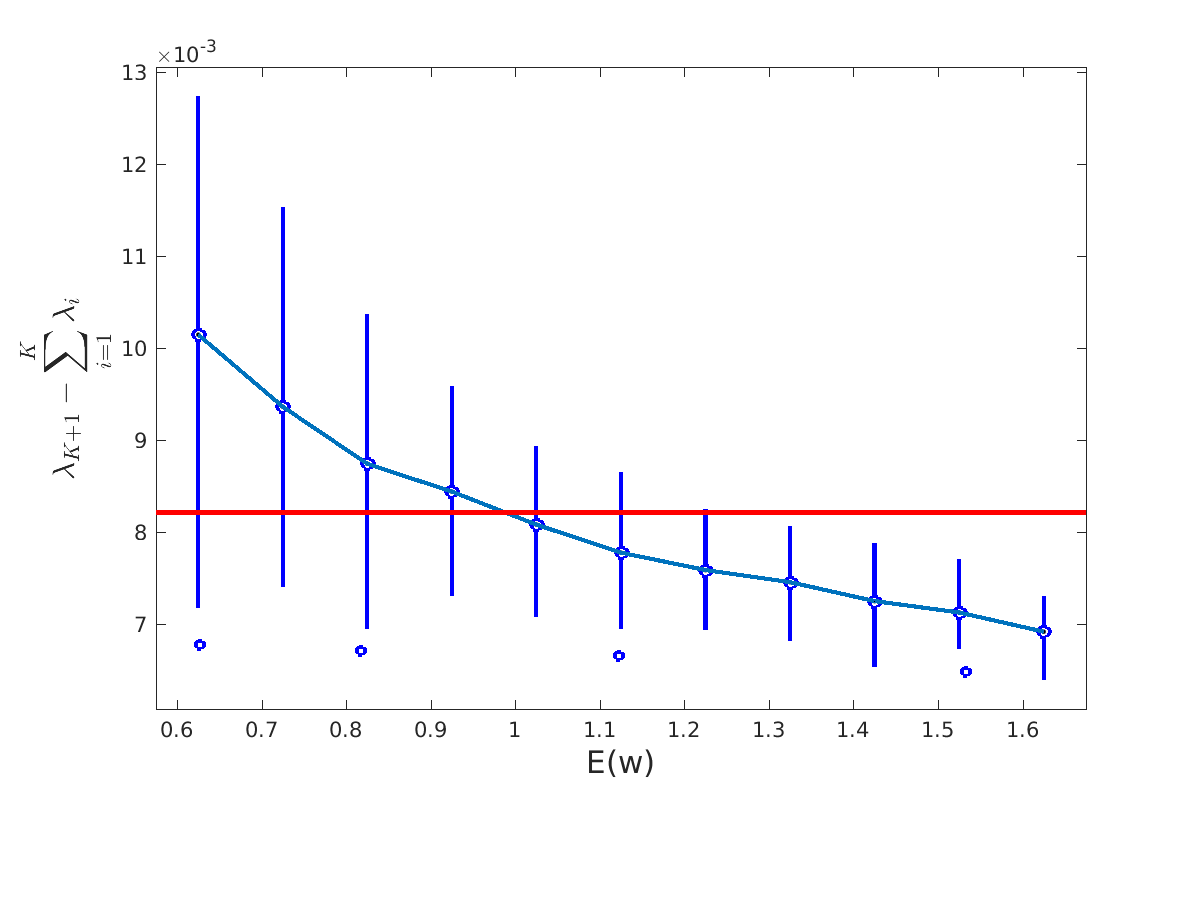}\\
\includegraphics[width=0.5\linewidth]{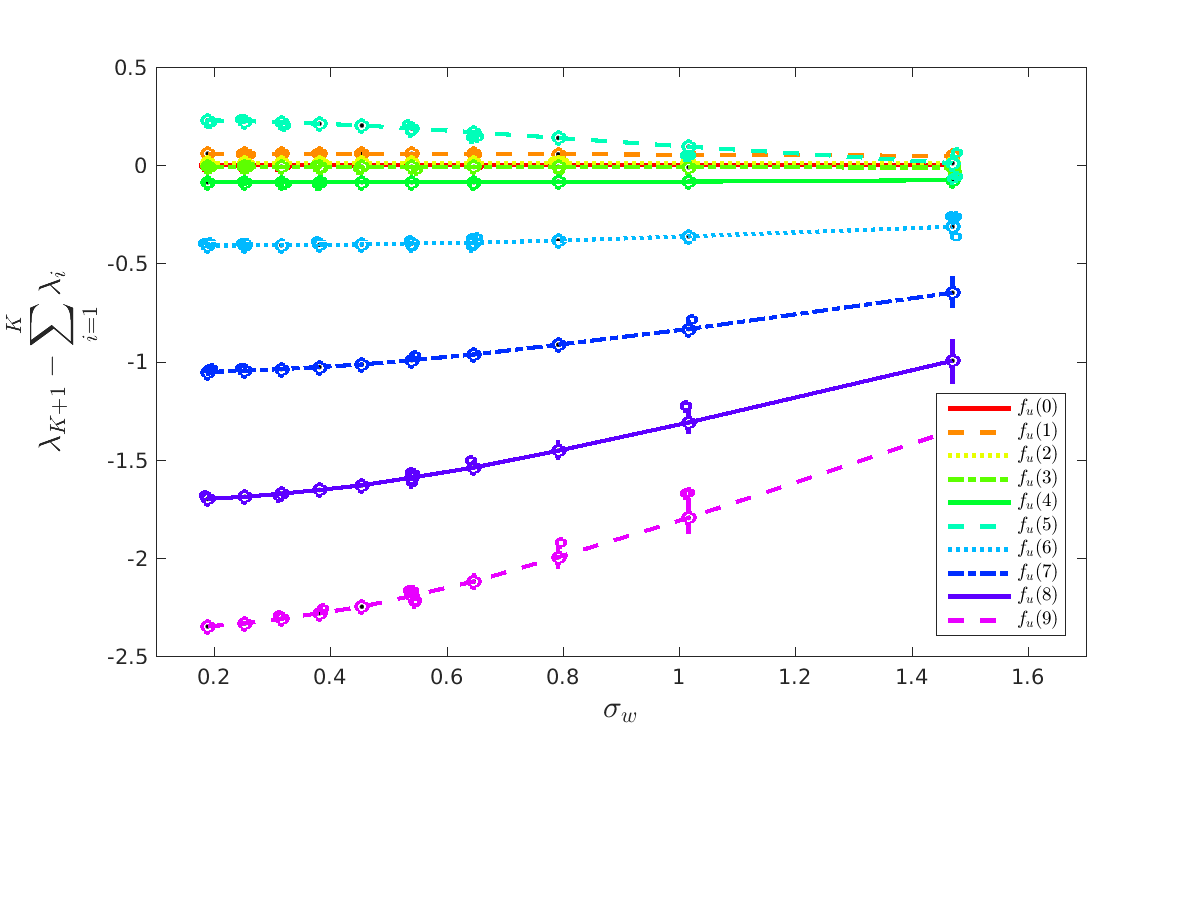}
\includegraphics[width=0.5\linewidth]{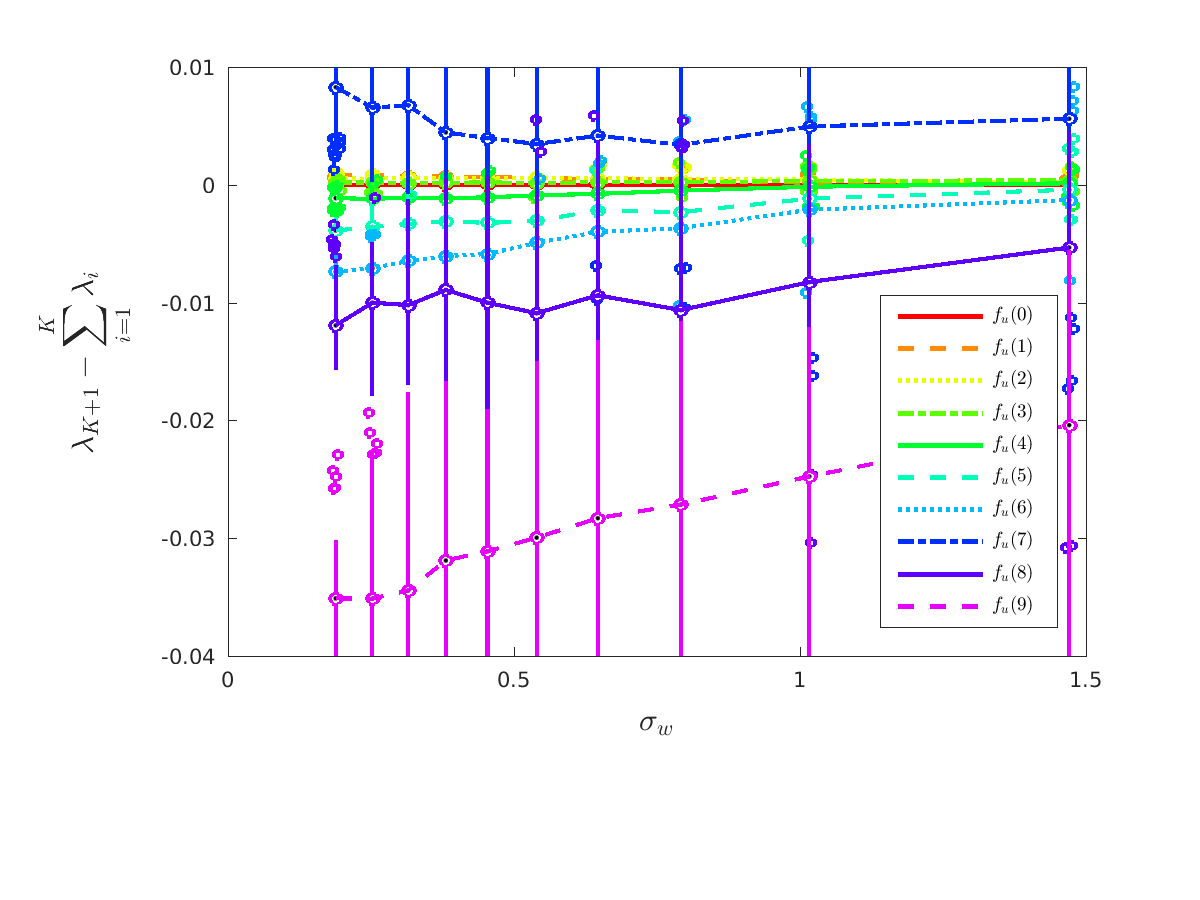}\\
\end{tabular}
\caption{\label{fig:5}Topleft: $IF(\lambda_2, w)$. Left column: synthetic datset. Right column: Facebook dataset. First row: $IF(f_l)$. Second row: $IF(f_u)$. Third row: breakdown point of $f_u$. Each boxplot is consist of 100 repetitions.}
\end{figure}

The synthetic dataset for testing the robustness of wCC's is generated with $n=2000$, and $\lambda_{1:K} = (0, 0.1, 0.11, 0.12, 0.13)$. In the synthetic dataset, because of the nature of the model,  the largest $f_l$ in the graph appears when $K=5$. In the Facebook dataset, through calculation ,we find that the largest $f_l$ appears when $K=7$. Therefore, we assume initially there are $5$ wCC's in the synthetic dataset and $7$ connected components in Facebook dataset. The results are shown in Figure \ref{fig:5}.\\

The synthetic dataset for testing the robustness of eigengap is generated with $n=800$ and $\lambda_{1:K} = (0, 0.1, 0.2, 0.3, 0.4)$. We call $f_e(i) = \lambda_{i+1} - \lambda_{i}$ the $i$th eigengap. Through calculation, we find that the largest eigengap is the $5$th in synthetic dataset, and $7$th in the Facebook dataset. \\
\begin{figure}[ht]
\begin{tabular}[b]{ccc}
\includegraphics[width=0.5\linewidth]{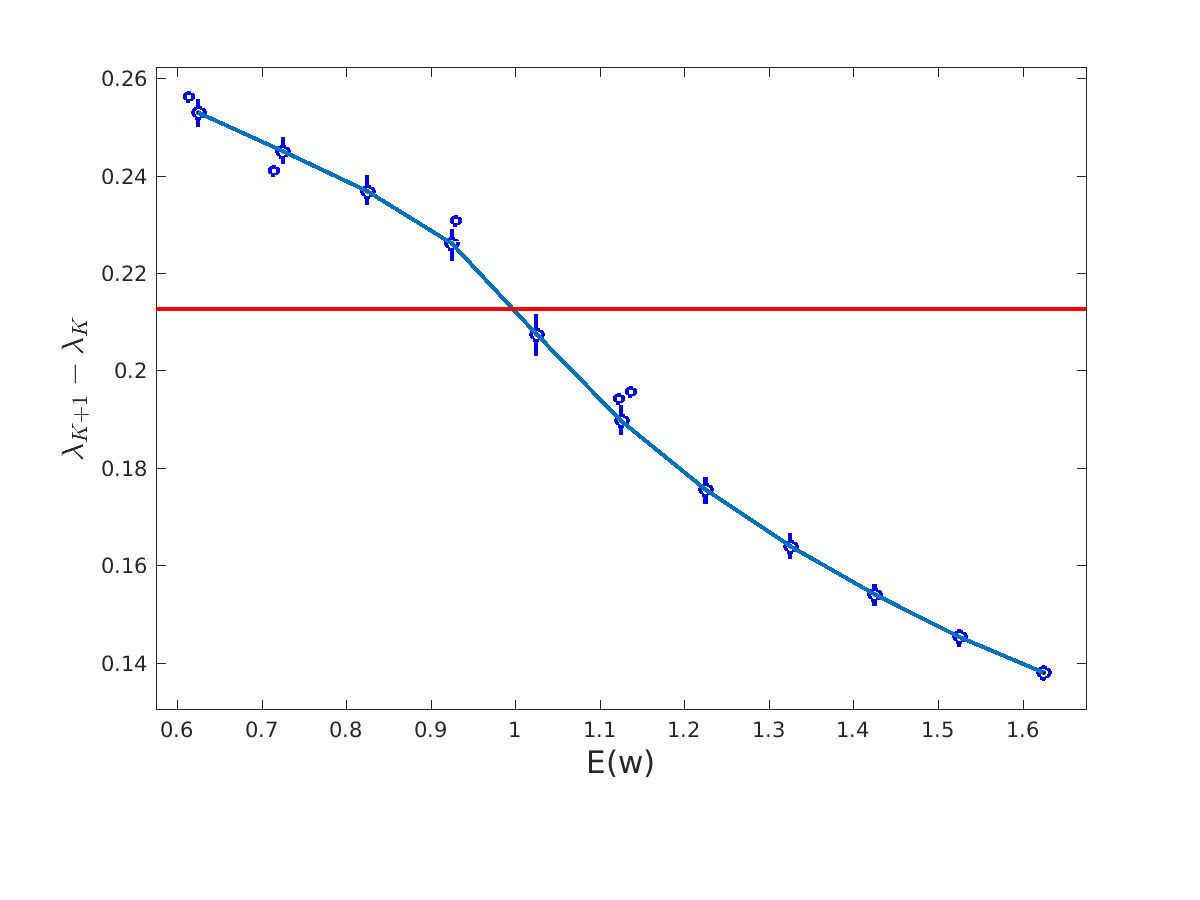}
\includegraphics[width=0.5\linewidth]{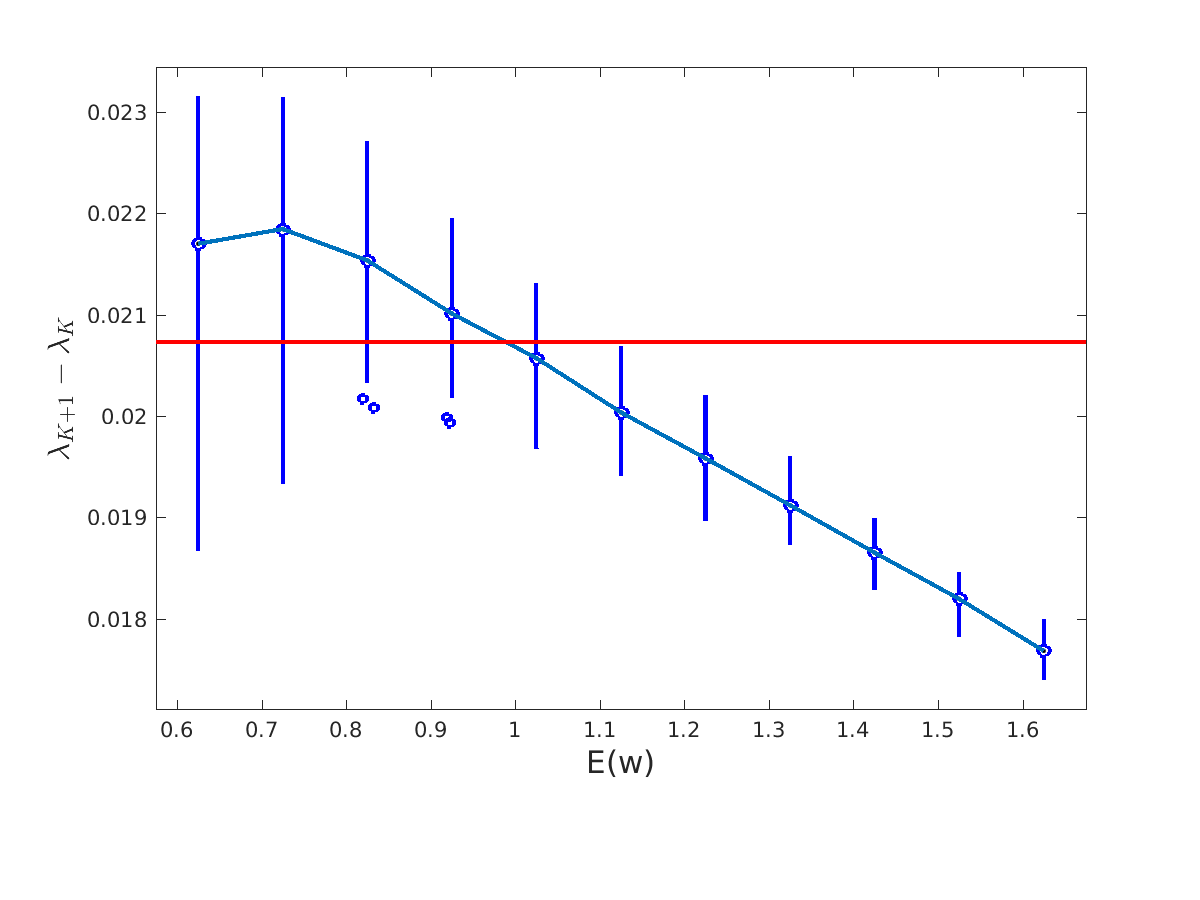}\\
\includegraphics[width=0.5\linewidth]{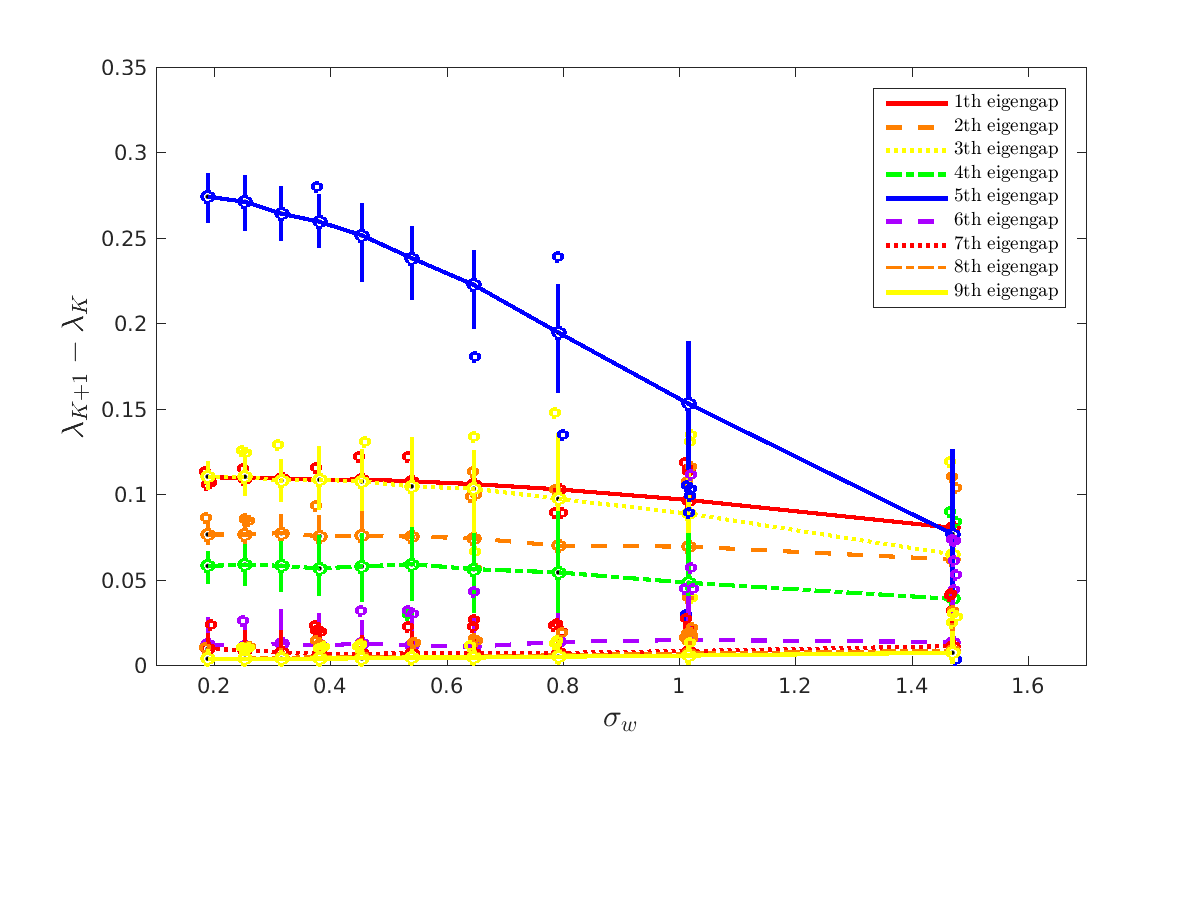}
\includegraphics[width=0.5\linewidth]{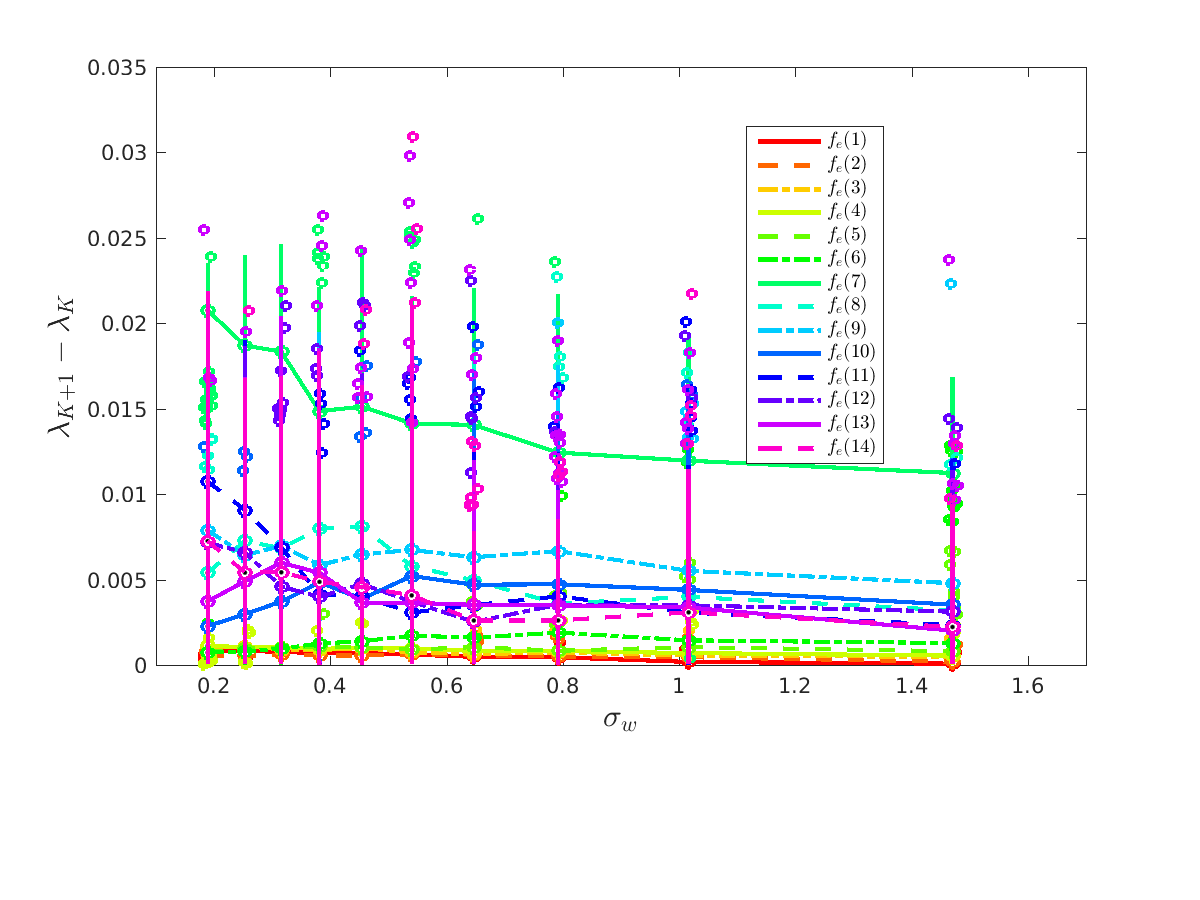}\\
\end{tabular}
\caption{\label{fig:gap}Left column: synthetic dataset. Right column: Facebook dataset. First row: $IF(f_e)$. Second row: breakdown point of $f_e$. Each boxplot is consist of 100 repetitions.}
\end{figure}

We observe that, $f_l$ increases while $f_u$ and $f_e$ decrease as $E(w)$ increases in both datasets, which indicates that, as the bad influence increases, there is greater connectivity between the connected components and the eigengap becomes less significant. \BP~is defined as the $\sigma_w$ where $f_u$, $f_l$ and $f_e$ is dominated with another $K$. We observe the \BP's for both the number of wCC's and eigengap. In the Facebook dataset, they seem to be robust to perturbation. Note that the Facebook dataset is consist of $10$ ego-networks while our experiments indicate that there are only $7$ robustness wCC's inside, meaning there are $3$ more connected users who are grouped to one wCC. \\

\section{Discussion} \label{sec:6}
This paper makes several contributions. Firstly, it provides an innovative way to perturb the network by assigning  the weights on the nodes and edges. The strength of perturbation can be well controlled and can be arbitrarily small. The topology of the graph is also preserved after the perturbation. Secondly, it extends the definitions of influence function and breakdown point to the graph properties. Although these measures are widely used in the robust statistics literature, using them on graph properties is the first time. Last but not the least, we are also able to probe the source of robustness by quantifying the influence of nodes on the robustness of graph properties, which provides a deeper insight into the problem. \\

Our perturbation framework also have its limitations. For example, no new edges or nodes can be added to the graphs through our perturbation methods, and this may not be natural evaluating the graph properties in some social networks. Moreover, the perturbation methods is not suitable for evaluating some graph properties, e.g. properties related to graph distances. These could be the area for future explorations.

\newpage
\bibliographystyle{plain}
\bibliography{ref}

\begin{thebibliography}{10}

\bibitem{ali2016comparison}
Waqar Ali, Anatol~E Wegner, Robert~E Gaunt, Charlotte~M Deane, and Gesine
  Reinert.
\newblock Comparison of large networks with sub-sampling strategies.
\newblock {\em Scientific reports}, 6:28955, 2016.

\bibitem{bhattacharyya2015subsampling}
Sharmodeep Bhattacharyya, Peter~J Bickel, et~al.
\newblock Subsampling bootstrap of count features of networks.
\newblock {\em The Annals of Statistics}, 43(6):2384--2411, 2015.

\bibitem{gfeller2005finding}
David Gfeller, Jean-C{\'e}dric Chappelier, and Paolo De~Los~Rios.
\newblock Finding instabilities in the community structure of complex networks.
\newblock {\em Physical Review E}, 72(5):056135, 2005.

\bibitem{Hampel:86}
Frank~R Hampel, Elvezio~M Ronchetti, Peter~J Rousseeuw, and Werner~A Stahel.
\newblock {\em Robust statistics: the approach based on influence functions},
  volume 114.
\newblock John Wiley \& Sons, 2011.

\bibitem{karrerLN:08}
B.~Karrer, E.~Levina, and M.E.J. Newman.
\newblock Robustness of community structure in networks.
\newblock {\em Physical Revieww E}, 77(4)::046119, 2008.

\bibitem{leskovec2012learning}
Jure Leskovec and Julian~J Mcauley.
\newblock Learning to discover social circles in ego networks.
\newblock In {\em Advances in neural information processing systems}, pages
  539--547, 2012.

\bibitem{magnus1988matrix}
Jan~R Magnus and Heinz Neudecker.
\newblock Matrix differential calculus with applications in statistics and
  econometrics.
\newblock {\em Wiley series in probability and mathematical statistics}, 1988.

\bibitem{MPentney:sdm07}
Marina Meil\u{a} and William Pentney.
\newblock Clustering by weighted cuts in directed graphs.
\newblock In Chid Apte, David Skillicorn, and Vipin Kumar, editors, {\em
  Proceedings of the SIAM Data Mining Conference, SDM}. SIAM, 2007.

\bibitem{nadler:06}
Boaz Nadler, Stephane Lafon, Ronald Coifman, and Ioannis Kevrekidis.
\newblock Diffusion maps, spectral clustering and eigenfunctions of
  fokker-planck operators.
\newblock In Y.~Weiss, B.~Sch\"{o}lkopf, and J.~Platt, editors, {\em Advances
  in Neural Information Processing Systems 18}, pages 955--962, Cambridge, MA,
  2006. MIT Press.

\bibitem{qin2013regularized}
Tai Qin and Karl Rohe.
\newblock Regularized spectral clustering under the degree-corrected stochastic
  blockmodel.
\newblock In {\em Advances in Neural Information Processing Systems}, pages
  3120--3128, 2013.

\bibitem{verzelen2015community}
Nicolas Verzelen, Ery Arias-Castro, et~al.
\newblock Community detection in sparse random networks.
\newblock {\em The Annals of Applied Probability}, 25(6):3465--3510, 2015.

\bibitem{WanM:isaim16}
Yali Wan and Marina Meila.
\newblock Benchmarking recovery theorems for the {DC-SBM}.
\newblock In {\em International Symposium on Artificial Intelligence and
  Mathematics (ISAIM)}, 2015.

\bibitem{MWan:nips15}
Yali Wan and Marina Meila.
\newblock A class of network models recoverable by spectral clustering.
\newblock In Daniel Lee and Masashi Sugiyama, editors, {\em Advances in Neural
  Information Processing Systems (NIPS)}, 2015.

\bibitem{WanM:nips16}
Yali Wan and Marina Meil\u{a}.
\newblock Graph clustering: block-models and model-free results.
\newblock In {\em Advances in Neural Information Processing Systems (NIPS)},
  2016.

\end{thebibliography}

\end{document}